\documentclass{aastex61}

\newcommand\aastex{AAS\TeX}

\received{}
\revised{}
\accepted{}
\submitjournal{ApJ}

\shorttitle{\aastex\ Dynamical effects}
\shortauthors{A. Maeder}
\begin{document}

\title{Dynamical effects of the scale invariance of the empty space:\\
the fall of dark matter ?}

%% you desire. In v6.1 you can include a footnote in the title.
%%
%% The \author command is the same as before except it now takes an optional
%% arguement which is the 16 digit ORCID. The syntax is:
%% \author[xxxx-xxxx-xxxx-xxxx]{Author Name}
%%
%% This will hyperlink the author name to the author's ORCID page. Note that
%% during compilation, LaTeX will do some limited checking of the format of
%% the ID to make sure it is valid.
%%
%% Use \affiliation for affiliation information. The old \affil is now aliased
%% to \affiliation. AASTeX v6.1 will automatically index these in the header.
%% When a duplicate is found its index will be the same as its previous entry.
%%
%%
%% The new \altaffiliation can be used to indicate some secondary information
%% such as fellowships. This command produces a non-numeric footnote that is
%% set away from the numeric \affiliation footnotes.  NOTE that if an
%% \altaffiliation command is used it must come BEFORE the \affiliation call,
%% right after the \author command, in order to place the footnotes in
%% the proper location.

\correspondingauthor{Andre Maeder}
\email{andre.maeder@unige.ch}

\author[0000-0002-0786-7307]{Andre Maeder}
\affiliation{Geneva Observatory \\
chemin des Maillettes, 51 \\
CH-1290 Sauverny, Switzerland}

\begin{abstract}
The   hypothesis of the scale invariance of the macroscopic  empty space, which intervenes through the
cosmological constant, 
 has led to  new  cosmological models. They
  show an  accelerated  cosmic  expansion after the initial stages and  satisfy  
several major cosmological tests \citep{Maeder17a}. 
 No unknown  particles are needed.
Developing   the weak field approximation, we find that  the 
here derived equation of motion corresponding to Newton's equation also contains a small outwards 
acceleration term. Its order of a magnitude  is about  $\sqrt{\varrho_{\mathrm{c}}/ \varrho}  \; \times$ Newton's gravity, 
 ($\varrho$ being  the mean density of the system  and $\varrho_{\mathrm{c}}$ the usual critical density).  The new term is thus 
 particularly significant for very low density systems.  

A modified virial theorem is derived and applied to clusters of galaxies. For  the Coma  and Abell 2029 clusters, the dynamical
masses are about a factor of 5 to 10 smaller than in the standard case. This tends to let  no room for dark matter in these clusters.
Then, the two-body problem is studied and an equation  corresponding to the Binet equation is obtained. It
implies some secular variations of the orbital parameters.
The results are applied to the rotation curve of the outer layers of the Milky Way. 
Starting backwards from  the present rotation curve, we  calculate the past evolution of the galactic rotation
and  find that, in the early stages, it was steep and Keplerian. Thus, the flat rotation curves of galaxies appears as an age
effect, a result consistent with recent observations of distant galaxies by \citet{Genzel17} and \citet{Lang17}.
Finally, in an Appendix we also study the long-standing problem of the increase with age of the vertical velocity dispersion in the Galaxy.
The observed increase appears to result from the new small acceleration term in the equation of the harmonic oscillator  describing stellar 
motions around the galactic plane. Thus, we tend to conclude that  neither the dark energy, nor the dark matter seem to be needed
in the proposed theoretical context. 

\end{abstract}


\section{Introduction: the context} \label{sec:intro}

The problem of the dark matter, noticeably  raised   decades ago  by the dynamical studies of clusters of galaxies 
and by the flat  rotation curves of galaxies,   is still resisting to explanations.
An impressive variety of  exotic particles   has
 been proposed in order to account for dark matter, see  recent reviews by  \citet{Bertone17} and by \citet{Swart17}.
  Simultaneously, theories of modified gravity like 
 the MOND theory \citep{Milgrom83} are not in arrears, as recently reviewed by \citet{Famaey12} and \citet{Kroupa12,Kroupa15}.
 In this interesting context, it may also be worth  to reconsider some basic physical invariances 
 of the gravitation theory.

 The group of invariances subtending  theories  plays a most fundamental role in physics \citep{Dirac73}.  The Maxwell equations 
 in absence of charge and currents are scale invariant, while the equations of General Relativity (GR) do not enjoy this additional property
  \citep{Bondi90}. We know that a general scale invariance of the physical laws  is prevented 
  by the presence of matter, which defines scales of mass, time and length  \citep{Feynman63}. 
  However, the  empty space at large
  scales  could have the property of scale invariance, since by definition there is nothing to define a scale.  
  The real space is never empty in the Universe, however the properties of the empty space intervene through 
  $\Lambda_{\mathrm{E}}$, the Einstein cosmological constant.
  It is true that the vacuum at the quantum
   level is not scale invariant, since 
   some units of mass, length and time can be defined on the basis of the Planck constant.
   However, the  large scale empty space differs by an enormous factor 
  from the quantum scales. Thus, alike
  we may apply Einstein's theory at large scales even if we cannot do it at the quantum level, we  may make the 
  scientifically acceptable  hypothesis that {\emph{the properties of the empty space represented by 
  $\Lambda_{\mathrm{E}}$ at macroscopic and  astronomical scales are scale invariant}}. 
  %The contribution of the empty space 
   %to the momentum-energy of the Universe is generally expressed by the cosmological constant.  
   In this
   work (see also \citet{Maeder17a}, hereafter called Paper I),  
   we are exploring further consequences of the above hypothesis.
   The MOND theory has been noted to have this property \citep{Milgrom09}, but since this is a classical theory, it is not contained in a cosmological model.
  
 The  consequences are far reaching, as shown  by the cosmological models in Paper I 
 which consistently include, through $\Lambda_{\mathrm{E}}$,
 the invariance of the empty space at macroscopic scales.
%%nicely confirm that  tiny amounts of matter very rapidly kill scale invariance. The problem is that 
%for a density parameter $\Omega_{\mathrm{m}}$ of about 0.3, the effects of scale invariance appear not to be completely killed.
These models clearly account for the acceleration of the cosmic expansion, without
calling for some unknown particles of any kinds. Several cosmological tests have been performed, 
they concern  the distance vs. redshift $z$ relation,
  the magnitude--redshift  $m - z$ diagram,
  % (Appendix B illustrates a comparison with a recent $m - z$ diagram), 
  the plot of the density parameters 
 $\Omega_{\mathrm{m}}$ vs. $\Omega_{\Lambda}$, the relations of
 the Hubble constant $H_0$ with  the age of the Universe and $\Omega_{\mathrm{m}}$, the past expansion rates $H(z)$
 vs. $z$ and  the transition from braking to acceleration, 
  and more recently  the  past  temperatures of the CMB vs. redshifts \citep{Maeder17b}.
 All these tests are impressively satisfactory and they open the possibility that the so-called dark energy may be
 an effect of the scale invariance  of the empty space at large scales.
  Therefore, it is scientifically  reasonable  to explore further consequences of the above 
 hypothesis to see whether at some stage it meets severe contradictions with the observations or whether it continues
 to show agreement. We now especially consider the dynamical evidences  of dark matter.
 
 As the  internal dynamics of clusters of galaxies and the rotation of galaxies are at the origin of the claim for the existence
 of dark matter, we  focus here on these dynamical problems.
 In Sect. 2, we study  the Newtonian  approximation of the geodesic equation  consistent with the above key hypothesis.
  In Sect. 3, we examine
 the dynamical or virial  masses of clusters of galaxies in the scale invariant context
  and apply our results to the Coma and Abell 2029 clusters. In Sect. 4,
 we  study the scale invariant two-body problem  and then discuss the outer  rotation curve of the Galaxy.
 The case of galaxies at significant redshifts is also considered. Sect. 5 gives brief conclusions. In an Appendix, we
 examine the age - velocity dispersion relation of stellar groups in the Galaxy, 
 in particular in the vertical direction where there
 is no consensus on the origin of the relation.

 \section{The Newtonian approximation of the scale  invariant field equations}  \label{Newton}
 
 \subsection{Brief recalls of cotensor analysis} 
 
 To express the scale invariance of the empty space
  intervening through $\Lambda_{\mathrm{E}}$ at large scales,
 we must consistently do it in a theoretical framework
  which permits     scale invariance  (but does not necessarily demand it !). General relativity does not offer 
 this possibility, however a framework, the cotensor analysis,
  that allows it has been worked out in details by \citet{Weyl23}, \citet{Eddi23}, \citet{Dirac73},
 \citet{Canu77}, (this was often in the context of the studies on variable $G$, but this is not what we do here).
 Short  summaries of cotensor analysis are given by  \citet{Dirac73} and in an Appendix by \citet{Canu77},
see alo \citet{BouvierM78}.
 %We also emphasize that  scale invariance was often considered as a general property of gravitation. 
  % The developments  are based in the mathematical context of the so-called cotensor analysis developed by the above authors. 
    In addition to the general covariance 
 of tensor analysis used in GR, cotensor analysis  also admits the possibility of scale  invariance  of the form
 \begin{equation}
ds' \, = \, \lambda(x^{\mu}) \, ds \, .
\label{lambda}
\end{equation}
\noindent
There,   $ ds'^2  \,= \, g'_{\mu \nu} dx'\,^{\mu} \, dx'\,^{\nu}$ is the line element in the GR framework with coordinates $ x' \,^{\mu} $
 and  $ds^2  =  g_{\mu \nu} dx^{\mu} \, dx^{\nu} $
is the line element in a new   more general framework, where we examine scale invariance. 
Parameter $\lambda(x^{\mu})$ is the scale factor connecting the two line elements.
If $\lambda(x^{\mu})=1$, the two frameworks are the same. In addition, we also make here a transformation of
coordinates from  $ x' \,^{\mu} $ to $ x \,^{\mu} $, because  we want to study simultaneously 
the effects of transformation of coordinates   as in GR together with the
effects of a change of scale.

When the various steps of the development of cotensorial analysis are followed, a general scale invariant 
field equation can  be written (see paper I).  
With respect to the usual field equation, it contains additional terms
depending only on $g_{\mu \nu}$ and  on $\kappa_{\nu}$, where
  \begin{equation}
\kappa_{\nu} \, = \, -\frac{\partial}{\partial x^{\nu}  } \, \ln \lambda \, .
\label{kappa}
\end{equation}
\noindent
The  term $\kappa_{\nu}$ is called the coefficient of metrical connection.
It is as a fundamental quantity  as are the $g_{\mu \nu}$ in GR.
The field equation writes \citep{Canu77}
\begin{eqnarray}
R'_{\mu \nu}   -  \frac{1}{2}  \ g_{\mu \nu} R' - \kappa_{\mu ;\nu}  - \kappa_{ \nu ;\mu}
 -2 \kappa_{\mu} \kappa_ {\nu}   
+ 2 g_{\mu \nu} \kappa^{ \alpha}_{;\alpha}
 - g_{\mu \nu}\kappa^{ \alpha} \kappa_{ \alpha}   =
-8 \pi G T_{\mu \nu} - \lambda^2 \Lambda_{\mathrm{E}}  \, g_{\mu \nu} \, .
\label{field}
\end{eqnarray}
\noindent
The terms with a prime are those of GR.
The gravitational constant $G$ is  a true constant and
 $\Lambda_{\mathrm{E}}$ the Einstein cosmological constant. The symbol '';'' means a derivative.
 The application of the general field equation to the empty space  has led in paper I to some relations between 
 the cosmological constant and the scale factor $\lambda$. The assumption is also made that the empty space is 
 homogeneous and isotropic, which implies  that scale factor $\lambda$ is only a function  of the cosmic time $t$. 
 The 1, 2, 3 components (the  three give the same result) and the 0 component of the above field equation  become
 respectively for the empty space \citep{MBouvier79,Maeder17a}
 \begin{equation}
 2 \, \frac{\dot{\kappa_0}}{c}-\kappa^2_0\, =\, -\lambda^2 \, \Lambda_{\mathrm{E}} \, , \quad \mathrm{and} \quad
 3 \, \kappa^2_0 \, = \,  \lambda^2 \, \Lambda_{\mathrm{E}} \, .
 \label{kk}
 \end{equation}
 \noindent
 The addition of these two equations gives $\frac{\dot{\kappa_0}}{c}= -\kappa^2_0$, the solution of which is
 \begin{equation}
 \kappa_0 \, = \, \frac{1}{c \, t} \, .
 \label{kzero}
 \end{equation}
 \noindent
 Here we keep the velocity of light $c$ in the equations in order to write the weak field equations with the appropriate units.  
 From Eq.(\ref{kappa}), one also has 
$ \kappa_0  = - \frac{\dot{\lambda}}{c \, \lambda}$. This expression together
 with Eqs.(\ref{kk})  leads to
% {\it{i.e.}} the energy-density of the empty space which permeates the Universe, 
%and the scale factor $\lambda$,
 \begin{eqnarray}
\  3 \, \frac{ \dot{\lambda}^2}{c^2 \,\lambda^2} \, =\, \lambda^2 \,\Lambda_{\mathrm{E}}  \,  
 \quad \mathrm{and} \quad 2 \, \frac{\ddot{\lambda}}{c^2 \,\lambda} - \frac{ \dot{\lambda}^2}{c^2 \,\lambda^2} \, =
\, \lambda^2 \,\Lambda_{\mathrm{E}}  \, ,
\label{diff1}
\end{eqnarray}  
\noindent
which give the fundamental relations between $\Lambda_{\mathrm{E}}$ and the scale factor $\lambda$.
We  see that if $\Lambda_{\mathrm{E}} = 0$,
the scale factor would be a constant, that is to say the scale invariant framework would be strictly identical to GR. 
The first of the above equations  leads to 
$ \lambda \, = A/t$, where $A$ is a constant. Taking 
$ \lambda=1 $ at  the present time  $t_0$, one has 
\begin{equation}
 \lambda \, = \, \frac{t_0}{t} \, .
\label{LE}
\end{equation}
\noindent
The origin of time $t$ depends on the cosmological models. For example,
the numerical models in Paper I  show that for $t_0=1$, the origin 
lies at $t_{\mathrm{in}}= 0.6694$ for a value of $\Omega_{\mathrm{m}}=0.30$. This means that the
variations of the scale factor $\lambda$ are small, being limited to a change from 1.4938 at the Big-Bang to 1.0
at present time. For $\Omega_{\mathrm{m}}=0$,  one has  $t_{\mathrm{in}}= 0$ and 
the variations of $\lambda$ would go from infinity to zero.
These examples  show that the presence of matter  rapidly reduces the cosmological  effects of  scale invariance, cf.  \citet{Feynman63}.

To study the dynamics of systems, we need an equation of motion. For that, we  may  derive the  geodesic equation
in the scale invariant framework in various ways \citep{MBouvier79}. Let us do it in  a straightforward way, starting 
from the Equivalence Principle  as expressed by \citet{Weinberg72}. At every point of the space--time, there is a local inertial system
$x'^{\alpha}$ such that the motion in GR may be described by
\begin{equation}
\frac{d^2 x'^{\alpha}}{ds'^2} \, = \, 0.
\end{equation}
\noindent
Let us develop this expression in the new framework (defined by $ds^2$),
\begin{equation}
\frac{d}{ds'} \, \left(\frac{\partial x'^{\alpha}}{\partial x^{\mu}} \frac{dx^{\mu}}{ds'}  \right) \, =
 \, \frac{d}{\lambda \, ds}\, \left(\frac{\partial x'^{\alpha}}{\partial x^{\mu}} \frac{dx^{\mu}}{\lambda ds}  \right) \,= \, 0 \, ,
 \label{geo1}
 \end{equation}
 
 \begin{equation}
 \frac{d^2 x^{\rho}}{ds^2}+ \frac{\partial^2 x'^{\alpha}}{\partial x^{\mu}\partial x^{\nu}}\frac{\partial x^{\rho}}{\partial x'^{\alpha}} 
  \frac{dx^{\mu}}{ ds}  \frac{dx^{\nu}}{ ds} + \kappa_{\nu}  \frac{dx^{\rho}}{ds}  \frac{dx^{\nu}}{ds} \, = \, 0 \, ,
  \label{geo2}
  \end{equation}
\noindent
 In cotensor analysis,  scale invariant derivatives of the first and 
 second order have been developed preserving  scale invariance. Other scale invariant quantities 
 are also defined, they are noted by a $^\ast$ \citep{Dirac73,Canu77}.
 The modified form of the Christoffel symbol
 $^{\ast}\Gamma^{\rho}_{\mu \,\nu}$  
  corresponds to the  first two  derivatives in the second term on the left of  Eq.(\ref{geo2})
 \begin{equation}
 ^{\ast}\Gamma^{\rho}_{\mu \,\nu} \, = \, 
 \frac{\partial^2 x'^{\alpha}}{\partial x^{\mu}\partial x^{\nu}}\frac{\partial x^{\rho}}{\partial x'^{\alpha}} \, .
 \label{Gstar}
 \end{equation}
 \noindent
 With (\ref{kappa}) and (\ref{Gstar}), we may write the equation of motion in the scale invariant framework,
  \begin{equation}
 \frac{d^2 x^{\rho}}{ds^2}+^{\ast}\Gamma^{\rho}_{\mu \,\nu}
  \frac{dx^{\mu}}{ ds}  \frac{dx^{\nu}}{ ds} + \kappa_{\nu}  \frac{dx^{\rho}}{ds}  \frac{dx^{\nu}}{ds} \, = \, 0 \, ,
  \label{geo3}
  \end{equation}
 \noindent
  The modified   Christoffel symbol  also writes (see relation (A5) by \citet{Canu77},
 (3.2)  by \citet{Dirac73} or (86.3) by \citet{Eddi23}),
 \begin{equation}
 ^{\ast}\Gamma^{\rho}_{\mu \,\nu} \, = \, \Gamma^{\rho}_{\mu \,\nu} \, -g^\rho_\mu \kappa_\nu - g^\rho_\nu \kappa_\mu
 +g_{\mu \nu} \, \kappa^\rho \, .
 \label{Gstar2}
 \end{equation}
 \noindent
 There,  $\Gamma^{\rho}_{\mu \nu}$ is the usual Christoffel symbol and the term
$\kappa_{\nu}$ is defined by (\ref{kappa}). 
 Quite generally, as shown by the field equation,
  the scale invariant  terms are given by the corresponding usual terms in GR, plus or minus some functions of
 the $g_{\mu \nu}$ and $\kappa_{\nu}$.
 %This also applies to   the Riemann-Christoffel tensor  $R^{\nu}_{\mu \lambda \rho}$,  to
 %its contracted form  $R^{\nu}_{\mu}$ and to the total curvature $R$.  Finally,  this is also true for the general scale invariant field equation,
% that is obtained by the developments of cotensor analysis (see paper I).
From relations (\ref{geo3}) and (\ref{Gstar2}), one has
\begin{equation}
 \frac{d^2 x^{\rho}}{ds^2}+\left(\Gamma^{\rho}_{\mu \,\nu} -g^\rho_\mu \kappa_\nu - g^\rho_\nu \kappa_\mu
 +g_{\mu \nu} \, \kappa^\rho \,\right) \,\frac{dx^{\mu}}{ ds}  \frac{dx^{\nu}}{ ds}
  + \kappa_{\nu}  \frac{dx^{\rho}}{ds}  \frac{dx^{\nu}}{ds} \, = \, 0 \, .
  \label{geo4}
  \end{equation}
 \noindent
 The third and the last terms simplify and one is left with the following geodesic equation,
\begin{equation}
\frac{du^{\rho}}{ds}+ \Gamma^{\rho}_{\mu \nu} u^{\mu} u^{\nu} -\kappa_{\mu}u^{\mu} u^{\rho}+ \kappa^{\rho} = 0 \, ,
\label{geod}
\end{equation}
\noindent
with the velocity  $u^{\mu} \, = \, dx^{\mu}/ds$. This equation  allows one to study the motion of astronomical bodies for various conditions.

%Finally, let us emphasize again that the above conclusions result from the hypothesis that the macroscopic empty space is scale
%invariant and that $\Lambda_{\mathrm{E}}$ is different from zero, {\it{i.e.}} that the energy density of the
%empty space is different from zero,  a current assumption in today's cosmology.

%The two expressions in (\ref{kk})
 %can also be written in term of $\lambda$. Respectively one 
% \begin{equation}
%\  3 \, \frac{ \dot{\lambda}^2}{\lambda^2} \, =\, \lambda^2 \,\Lambda_{\mathrm{E}}  \, ,
%\label{diff1}  
%\end{eqnarray}
%\noindent
%while the 1, 2,  3 components give the same following relation,
%\begin{eqnarray}
 %2 \, \frac{\ddot{\lambda}}{\lambda} - \frac{ \dot{\lambda}^2}{\lambda^2} \, =
%\, \lambda^2 \,\Lambda_{\mathrm{E}}  \, .
%\label{diff2}
%\end{eqnarray}
%\noindent
%or  in equivalent forms,
%\begin{eqnarray}
%\frac{\ddot{\lambda}}{\lambda} \, = \,  2 \, \frac{ \dot{\lambda}^2}{\lambda^2} \, , \quad
% \quad \mathrm{and} \quad \frac{\ddot{\lambda}}{\lambda} -\frac{ \dot{\lambda}^2}{\lambda^2} \,
 % = \, \frac{\lambda^2 \,\Lambda_{\mathrm{E}}}{3} \, .
%\label{diff2}
%\end{eqnarray}

\subsection{The weak field approximation} \label{sub:weak}

The Robertson-Walker metric was used to derive the cosmological equations from the general field equations in paper I.
These equations were greatly simplified 
thanks to relations (\ref{diff1}), that allow us to express $\Lambda_{\mathrm{E}}$.
 Compared to the usual standard equations of cosmology, they only  contain one
additional term representing an acceleration opposed to gravity, cf. Eq.(\ref{E3}) below.  
In view  of Eq.(\ref{LE}),  the effects due to the evolution over  a long period of time
 are expected to be the largest ones. The effect not depending on a time evolution are in principle  the same as in GR.
%differing only slightly from the Minkowski metric. 
%In this context, it may be recalled that in GR the Minkowski
%metric is formally incompatible with a non-zero cosmological constant. 
 %However, we have shown that in the scale invariant context
%the Minkowski metric is quite compatible with  the presence of  $\Lambda_{\mathrm{E}}$ satisfying Eq.  (\ref{diff1}) and ({\ref{diff2}).

Now, let us consider a test particle  in the  weak field   of a potential $\Phi$ created by  a central mass point. 
%For low velocities compared to the light velocity,    $u^{\mu}$ may be simply   written   $u^{\mu} \, \approx \, dx^{\mu}/ds$. 
We now develop this non-relativistic approximation, with $v/c  \ll 1$, of the  geodesic equation (\ref{geod}), 
 which in the classical framework
 would lead to Newton's equation. 
 The adopted metric only slightly deviates from the Minkowski metric,
\begin{equation}
 g_{i \, i} = -1\, , \; \mathrm{for} \;  i=1, 2, 3 \quad  \mathrm{and} \quad g_{00}=1+ (2\Phi/c^2) \, . 
 \end{equation}
 \noindent
 This implies that   the only non--zero component of the Christoffel symbols is \citep{Tolman34}
 \begin{equation}
 \Gamma^i_{00} \,=\, \frac{1}{2} \, \frac{\partial g_{0 0 }}{\partial x^i}\, =
  \,\frac{1}{2} \, \frac{\partial \left(1+ (2\Phi/c^2) \right)}{\partial x^i} \, =
 \frac{1}{c^2} \frac{\partial \Phi}{\partial x^i} \,.
 \end{equation}
 \noindent
 We also have $ds \approx c dt$ and the velocities are $u^i \approx \frac{v^i}{c} = \frac{dx^i}{c dt}$  and $u^0 \approx  1$.
 The only non--zero component of the coefficient of metrical connection $\kappa_{\nu}$ is $\kappa_0$. Thus, the last
 term in Eq.(\ref{geod}) vanishes.
% The scale factor $\lambda$ is that in $1/t$ derived above from Eq. (\ref{diff1}). 
 %The situation may be different near a black whole where the metric would be different. 
 In the Newtonian--like approximation of the equation of motion
 we have, 
 \begin{equation}
 \frac{1}{c^2} \frac{dv^i}{dt}+\frac{1}{c^2}  \frac{\partial \Phi}{\partial x^i } - \kappa_0  \frac{v^i}{c} \, = \, 0 \, .
 \label{geod2}
 \end{equation}
 \noindent
 In the cosmological models of paper I, we have put $c=1$, while  this is not the case here. Also, since  $\kappa_0$
 is a function of time (cf. Eq.\ref{kzero}), we define in order to avoid any ambiguity hereafter,
  \begin{equation}
   \kappa(t) \, \equiv \, c \,\kappa_0\,=\,1/t  \, .
   \label{kdef}
   \end{equation}
   \noindent
    Thus, one has 
 \begin{equation}
  \frac{dv^i}{dt}+  \frac{\partial \Phi}{\partial x^i } - \kappa(t)  v^i \, = \, 0 \, ,
 \label{geod22}
 \end{equation}
 \noindent
%where $\kappa(t) = c \,\kappa_0\,=\,1/t $  is the inverse of a time, {\it{i.e.}} in  $s^{-1}$ or  km s$^{-1}$ Mpc$^{-1}$.  Hereafter,
%we shall  write $\kappa (t)$ instead of $c \,\kappa_0$, in order to avoid any misunderstanding.  
 We need to express the  appropriate potential $\Phi$. In the framework of GR, we  would consider
  a  central mass point $M'$ and examine the situation at a distance $r'$ in a spherically symmetric system
  with a  potential  $\Phi'=- G M'/r'$.  In the scale invariant system, 
 from Eq.(\ref{lambda})  we have the correspondence $r'  = \lambda r$.  For the density, 
 it is    $\rho = \rho' \, \lambda^2$ according to Eq.({11})   in  Paper I. 
 Thus, $\frac{M}{r^3} \, = \, \frac{M'}{r'^3} \, \lambda^2$  and 
 the relation between the Einsteinian mass $M'$ and the scale invariant one is,
 \begin{equation}
 M' \,= \, \lambda \, M \, .
 \label{mass}
 \end{equation}
 \noindent
 The number of particles forming an object does evidently not change with time.
 Expression (\ref{mass}) is quite interesting: since  the mass  is changing like   the length  is doing, this means that
  the   curvature of space-time  (or  the gravitational potential) associated to a massive object is a scale invariant quantity,
 \begin{equation}
 \Phi' \, =- \frac{GM'}{r' } \, = \,- \frac{GM}{r } =\, \Phi \, \, ,
 \label{Phi}
 \end{equation}
 \noindent
  being the same in the GR and scale invariant frameworks.
 Eq.(\ref{geod2}) applies to each of the $i$--components. In Cartesian coordinates we may write
 \begin{equation}
 \frac {d^2  x^{i}}{dt^2} \, = \, - \frac{G \, M}{r^2} \, \frac{x^{i}}{r}   + \, \kappa(t)\,  \frac{dx^{i}}{dt} \, .
 \label{cartes}
 \end{equation}
 \noindent
 In spherical coordinates, we can write the vectorial form of the 
   equation  of motion  
 \begin{equation}
 \frac {d^2 \bold{r}}{dt^2} \, = \, - \frac{G \, M}{r^2} \, \frac{\bold{r}}{r}   + \, \kappa(t)\,  \frac{d\bold{r}}{dt} \, 
 %\quad \mathrm{or} \; \; 
 %\frac {d^2 \bold{r}}{dt^2} \, = \, - \frac{G \, M'}{\lambda \,r^2} \, \frac{\bold{r}}{r}   + \, \kappa_0 (t)\,  \frac{d\bold{r}}{dt} \, 
\label{Nvec}
\end{equation}
\noindent
We recognize  the Newton's law plus an additional term opposed to gravity.
This expression means that in addition to the curvature of space associated to a mass element  a particle may experience some 
outwards acceleration associated to the non-constancy   of the scale factor $\lambda$. This additional term  is generally very small, 
as discussed  in Sect. \ref{magn}.

%We note that in a short interval of time  around  present  time $t_0$,  we may take $\kappa_0$ as constant
%and thus  $dv  = \kappa_0  \, dr$, which gives 
%\begin{equation}
%v-v_0 \, = \, \kappa_0 \, (r-r_0) \, .
%\label{h}
%\end{equation}
%\noindent
%It means that locally the effect of the additional term is rather similar to the Hubble law.
%This is evidently a   very different view from the classical
%picture, where there is no effect of the Hubble expansion within a gravitationally bound system \citep{Misner73}.
%(In this connection, we may remark that the limit where Hubble law stops working has never been clearly defined).
%This  local expansion   is consistent the detailed study of the  two--body 
%problem in the scale invariant context  studied in Sect. \ref{twobody}.

 We have to take the proper units of time  in Eq.(\ref{Nvec}). In current units, the present age $t_0$ of the Universe is
 13.8 Gyr \citep{Frie08} or  $4.355 \cdot  10^{17}$ s. (This is an observed age value independent on cosmological models, resting
  essentially  on  the rather uncertain ages of globular clusters, see \citet{Catelan17} for a review. It is clear however that the relation between the age and  a parameter
  like $H_0$ depends on the cosmological models, see below.)
  The inverse of the above age  is $2.295 \cdot 10^{-18}$ s$^{-1}$ 
 or 70.85 km s$^{-1}$ Mpc$^{-1}$. Thus, the empirical value of $\kappa(t_0)= 1/t_0$ is a quantity very close to the
  current value of the Hubble constant $H_0$,   which lies between
 73  and 67 km s$^{-1}$ Mpc$^{-1}$ \citep{ChenRatra11,Aubourg15,Riess16,Chen16}.
 We may write the relation between the Hubble constant and the age  $t_0$  of the Universe 
 in some chosen cosmological models  as follows
 \begin{equation}
 H_0 \, = \,  \xi \, \frac{1}{t_0} \, ,
 \label{xi}
 \end{equation}
 which may be written for other times with appropriate $\xi$.
 The numerical factor $\xi$ depends on the cosmological model. For scale invariant  models with $\Omega_k = 0$ and
 values of  $\Omega_{\mathrm{m}}=0.10, 0.20, 0.25, 0.30, 0.40$ 
we have $\xi \, = \, 1.191, 1.038, 0.987, 0.945, 0.878$ respectively (cf. column 7 in Table 1 of paper I).

Some developments of the weak field 
approximation  were already performed  \citep{MBouvier79}. However, at that time $\kappa(t)$  
was identified with the Hubble constant. Although  the numerical values are very close to each other, 
there is an important  physical difference between the two.  The Hubble constant % $(\dot{R}/R)_0$
depends on the cosmological models, with $H_0 = (\dot{R}/R)_0$ being the result of the  evolution of the Universe
for the appropriate  parameters
 $\Omega_{\mathrm{m}}$ and  $\Omega_{k}$. This is physically different from the properties of the scale factor $\lambda$, 
 which results from Eqs. (\ref{diff1}).

Below, we shall carefully  explore some first consequences of the above law of mechanics (\ref{Nvec}).
% such as the virial theorem and its application to clusters of galaxies, the two--body problem and  the rotation  curves of galaxies.
 Observations, rather than dogmas,
 will tell us whether the above modified Newton's law  should be supported or  rejected.

\subsection{The order of magnitude of the new term}  \label{magn}

Let us  estimate numerically the relative importance of the additional acceleration term with respect to the Newtonian attraction
at the present  time $t_0$. We
consider  a test particle orbiting with a circular velocity $v$ at a distance $r$ of a point mass $M$. The ratio $x$  of 
the outwards acceleration term resulting from the scale invariance of the empty space with respect to the Newtonian inwards
attraction term in Eq.(\ref{Nvec}) is given by
\begin{equation}
x \, = \, \frac{ v \, r^2}{G \, M \, t_0} \, .
\end{equation}
\noindent 
%The present time $t_0$ is of the order of  the inverse of the Hubble constant $H_0$ and we write  
%\begin{equation}
%t_0 \,= \, \xi/H_0 \,  .
%\label{xi}
%\end{equation}
%which may also be written for other values of  $t$ and $H$.oindent
%The numerical factor $\xi$ depends on the cosmological model. For flat  models with $\Omega_k = 0$ and
% values of  $\Omega_{\mathrm{m}}=0.10, 0.20, 0.25, 0.30, 0.40$ 
%we have $\xi \, = \, 1.191, 1.038, 0.987, 0.945, 0.878$ respectively (column 6 in Table 1 of paper I).
We may now relate the present time $t_0$ to $H_0$ by expression (\ref{xi}) with the appropriate value of $\xi$, 
recalling that for the most realistic values of the density 
parameters $\Omega_{\mathrm{m}}$ the value of $\xi$ is of the order of unity.
In turn, $H_0$ may be related to the the critical density of the Universe at the present time. 
We have seen in Paper I that the true critical density $\varrho^*_{\mathrm{c}}$  corresponding to $k=0$ 
in scale invariant models is  given by 
(cf. Eq.(39) of paper I)
\begin{equation}
\varrho^*_{\mathrm{c}} \, = \, \frac{3}{8 \pi \, G}\left(H^2_0 - 2 \frac{H_0}{t} \right) \, = 
 \, \varrho_{\mathrm{c}}\left(1 - \frac{2}{3}  \; \frac{8 \pi \, G}{H_0 \, t}\right) \, .
\label{rhostar}
\end{equation}
\noindent
There, 
\begin{equation}
\varrho_{\mathrm{c}} \,=\, \frac{3 \,H^2_0}{8 \, \pi \, G}  
\label{rhoc}
\end{equation}
\noindent
is the standard critical density in Friedman's models. These densities are usually considered at the present time $t_0$, but the
above forms could also be used for other  epochs with the appropriate $H$--  and $t$--values.
Now, the above  $x$--ratio can be written in term of the standard density $\varrho_{\mathrm{c}}$ 
(the use of $\varrho^*_{\mathrm{c}} $   brings  other expressions with no particular interest for the numerical estimates below),
\begin{equation}
x\, = \, \frac{H_0 \,  v \, r^2}{\xi \, G \, M } \,= 
\, \frac{\sqrt{2}}{\xi}\,\left( {\frac{\varrho_{\mathrm{c}}}{\varrho }} \,\frac{v^2}{(GM/r)}\right)^{1/2} \, .
\end{equation}
\noindent
There, $\rho $ is the mean density associated to the mass $M$ within the radius $r$ considered. 
(At a time $t$ different from the present one, the corresponding values of the parameters 
 need to be taken.)
We will see, when studying the energy properties in Sect. \ref{vvir}, that the ratio $\frac{v^2}{GM/r}$ 
 is not necessarily always equal to unity.
%The velocity $v$ is the total velocity (radial + tangential). 
 As  the dynamical evolution of a system proceeds, 
 %the circular component  keeps constant as shown by Eq.(\ref{velocity}) below, while radially 
 the additional   acceleration term in Eq.(\ref{Nvec}) may introduce progressive deviations
 from the classical relation $v^2 \simeq G\,M/r$.
  According to    Sect. \ref{clusters},  the above ratio   $\frac{v^2}{GM/r}$   significantly differs from unity 
  only for systems with a density within less than about 3 order of magnitude 
  from the critical density $\varrho_{\mathrm{c}}$. Thus, we  write 
 \begin{equation}
 x \, \geq  \,   \frac{\sqrt{2}}{\xi}\,\left( {\frac{\varrho_{\mathrm{c}}}{\varrho }} \right)^{1/2} \, .
 \label{x}
 \end{equation}
 \noindent
  For systems  with $\varrho > 10^3 \, \varrho_{\mathrm{c}}$, we may consider the equality in the above expression.
% Only for systems much denser than critical, we may consider the equality in the above expression.}}
 The  ratio $x$ is thus mainly given by the ratio of the critical density  to
 the average density of the dynamical system considered. We see that the dynamical effects of the scale 
 invariance of the empty space  are  particularly significant in  systems  of very low density, such as clusters of galaxies 
 and possibly galaxies.
 
% The above relation (\ref{x}) also means that the gravitational interaction dominates the evolution of the astronomical systems with
 %a mean density superior to the critical density $\varrho_{\mathrm{c}}$.
 The acceleration term would dominate over gravitation  ($x > 1$)
  only for systems with $\varrho$ smaller than about $ 2  \, \varrho_{\mathrm{c}}$.
  The only such system known
 is the Universe, which presently shows some cosmic acceleration. The matter density and the critical density have 
 different time dependences. The matter density evolves according to the conservation law given by Eq.(61) in paper I, while the critical density 
 varies like $H^2$, (the variation of $H(z)$ are given 
in Table 2 of paper I for two useful models). The result is that
the acceleration term dominates over braking   only after a transition phase,
which is located near $z = 0.75$ for models with  $\Omega_{\mathrm{k}} = 0$ and $\Omega_{\mathrm{m}} = 0.30$ \citep{Maeder17a}.

\subsection{Consistency of the modified Newton equation and the cosmological equations}  \label{cons}

The scale invariant cosmological models depend on the usual density--parameters $\Omega_{\mathrm{m}}$ and
$ \Omega_{\mathrm{k}}$,  which now satisfy a relation of the form (see Eq.(45) in paper I),
\begin {equation}
\Omega_{\mathrm{m}} \, + \, \Omega_{\mathrm{k}} \, +  \Omega_{\lambda} = \, 1  \, ,
\quad \mathrm{with} \quad  \quad\Omega_{\lambda}  \equiv  \frac{2}{ H \, t} \, .
\label{Omegapr}
\end{equation}
\noindent 
For  models with $k=0$ supported by the observations of the CMB radiation \citep{deBern00,Benn03}, expansion implies $H>0$ and
thus $\Omega_{\lambda}  > 0$ with $\Omega_{\mathrm{m}} < 1$. As in standard cosmology,
from the two fundamental cosmological equations, a third one may  be derived  by elimination of
 the terms depending on the space curvature (paper I), it is
 \begin{equation}
-\frac{4 \, \pi G}{3} \, (3p +\varrho)  =  \frac{\ddot{R}}{R} + \frac{\dot{R} \dot{\lambda}}{R \lambda}  \, .
\label{E3}
\end{equation}
\noindent 
Terms $p$ and $\varrho$ are the pressure and density in the scale invariant system. $R(t)$ is the expansion function.
Taking $p=0$ and considering that the 
density $\varrho$ is the average density in a sphere of radius $R$ and central mass $M$. We get
\begin{equation}
\ddot{R} \, = \, - \frac{G \, M}{R^2} - \frac{\dot{\lambda}}{\lambda} \, \dot{R} \, ,
\end{equation}
\noindent
which compares  with  Eq.(\ref{Nvec}). This shows the consistency of the above modified Newton equation 
with the  scale invariant cosmological equations in their limit.

%This being clarified, let us examine the effects of the additional term in (\ref{Nvec}). 
Let us now consider the case of the empty space.  In  the Newtonian framework, a test particle would have a constant velocity
with $dv/dt=0$.
In the scale invariant case, it would experience a slow acceleration. 
From  the  additional term in Eq.(\ref{Nvec})  we have  $\frac{dv}{dt} \, = \, \frac{v}{t} $  and  thus
$v \, = \, a \, t \, ,  \quad \mathrm{and} \; \; r -r_0 \,= \, a \, (t^2 -t^2_0)$.
This is  quite consistent with the results of paper I, which show that the expansion function $R(t)$ 
of an empty universe would  behave like  $R(t)  \, \sim \, t^2$
in the scale invariant cosmology, while the empty Friedman model
would expand like  $R(t) \sim  t$.

\section{Dynamics of the clusters of galaxies}   \label{clusters}

Clusters of galaxies play an essential role in the determinations of the cosmological parameters \citep{Allen11}.
Their distribution as a function of redshifts depend on the geometry of the universe and on the growth of density 
fluctuations, which both in turn depend on $\Omega_{\mathrm{m}}$ and $\Omega_{\Lambda}$ \citep{Frie08}.
The determination of the virial masses was the first applied method to obtain the mass of the clusters of galaxies
\citep{Karachantsev66,Rood72,Bahcall74,Abell77,Blindert04,Proctor15}. It was soon evident that the estimated virial masses were much
too large compared to the visible  mass in galaxies. %This was often phrased in different equivalent ways, e.g.
 %the observed velocities of the galaxies  are too large for them to remain bound in clusters, or the visible mass is
  %very insufficient  to explain the 
%observations of the rapid motions of galaxies in clusters. 
%With the problem of the flat rotation of galaxies (Sect. xx), this was 
%a strong argument in favor of  a large amount of hidden mass or dark matter.

Specifically, we may consider that the stellar mass fraction $f_{*}=M_{\mathrm{star}}/M_{\mathrm{tot}}$ with respect 
to the total gravitational mass is of the order
 \begin{equation}
 f_{*} \,   \simeq \, \frac{(M/L)_{\mathrm{ref}}}{(M_{\mathrm{tot}}/L)} \, .
 \end{equation}
 \noindent
 There,  $(M/L)_{\mathrm{ref}}$ is the reference mass--luminosity ratio for a typical stellar population in galaxies
 and  $(M_{\mathrm{tot}}/L)$ is the total gravitational mass--luminosity  ratio  determined for  clusters of galaxies.
 $M_{\mathrm{tot}}$  is   the total gravitational mass, also called the dynamical mass  or virial mass  as it is determined on 
the basis of the  virial theorem in standard Newtonian dynamics.  
 The optical luminosity of galaxies originate mainly from stars, while  the total gravitational mass is that 
 of the baryons (stars, gas)  and dark matter.
  From 600 groups and clusters of galaxies studied in various color bands, 
 \citet{Proctor15} supported values of $(M_{\mathrm{tot}}/L) = (300 - 500) \;  (M/L)_{\odot} $
  for clusters with masses between $10^{14}$ and $10^{15}$
 M$_{\odot}$.  This well compares  to most  data by previous authors.
 % e.g. \citet{Bahcall74}, who found a ratio corresponding to 
 %490 for $H_0 =70$  km s$^{-1}$ Mpc$^{-1}$.
 For a typical  stellar value of $(M/L)_{\mathrm{ref}}=10 \, (M/L)_{\odot}$, 
 one obtains $f_{*}= 0.02$ to 0.033, a value well supported by the recent works mentioned  below.

 Recent results from optical and X-ray observations of clusters of galaxies 
 \citep{Andreon10,Lin12,Leauthaud12,Gonzalez13,Shan15,Ge16,Chiu16}
confirm that the stellar mass fraction $f_{*}=M_{\mathrm{star}}/M_{\mathrm{tot}}$ is quite small.
 Moreover  $f_{*}$  significantly 
 decreases with increasing total mass (virial),
typically from about 0.04 to less than 0.015 for cluster with masses from  $10^{14}$ to $10^{15}$ M$_{\odot}$.
 Measurements of the X--ray emitting  gas provide estimates of  the gas  fraction
  $f_{\mathrm{gas}}=M_{\mathrm{gas}}/M_{\mathrm{tot}}$, which largely dominates with respect
 to the stellar  mass fraction $f_{*}$. 
 %There is also an impressive increase of the  X--ray luminosities with  the  cluster masses    \citep{Stanek06}. 
 Results by the above authors  also show that the gas mass fraction  $f_{\mathrm{gas}}$
  increases from about  0.08 to 0.15 over the above mentioned mass interval.
   % \citep{Andreon10,Lin12,Leauthaud12,Gonzalez13,Ge16,Chiu16}.
 Clearly, most  baryons  reside in the hot gas. The  baryons fraction 
 $f_{\mathrm{bar}} = \frac{M_{\mathrm{star}}+M_{\mathrm{gas}}}{M_{\mathrm{tot}}}$, due to the opposite trends of the 
 stellar and gas components,  appear
  to  be nearly constant with cluster mass (around 0.12 to 0.15), {\it{e.g.}}  \citet{Gonzalez13}. 
   However,  this baryon fraction obtained  from the addition 
 of the  stellar and gas  components  appears,
  according to the above  authors,  slightly  lower than the cosmic WMAP-7yr  and Planck-2013
    value of  $f_{\mathrm{bar}}=0.17 $  and 0.157 respectively. 
  Whether this slight difference comes from uncertainties of the virial masses is a possibility \citep{Chiu16}. 
  The major  fact is that  the above baryon fraction   $f_{\mathrm{bar}}$ is much lower than 1, by about a factor of 6.
 This  is considered as a  strong  evidence in favor of the existence of dark matter.
  
   Now, 
  we may wonder whether a part of the difference between the total gravitational mass and the baryonic mass
  could  possibly  originate from the scale invariant dynamics.

%Gravitational lensing  provide another independent estimate of dark matter in clusters of galaxies \citep{Leauthaud10,vonderLinden14}.
%These various methods globally lead to the view that the intracluster gas is equivalent to a small multiple of the mass in galaxies

%Clusters of galaxies are  the largest  and the lowest density objects in the Universe. They are thus very  favorable 
%for testing the possible effects of the additional term in the scale invariant form of  the Newton equation.
 
\subsection{Scale invariance and the virial masses}  \label{vvir}

We may not directly apply the virial theorem in the context of the scale invariant theory, because of the additional term in Eq.(\ref{Nvec}),
which  produces the expansion of a gravitational system (see Sect. \ref{twobody}).
 The additional radial
outwards acceleration term in  Eq.(\ref{Nvec}) may influence the relation between the motions  and
 the present mass in a cluster of galaxies \citep{Maeder78}. We consider in a simplified way a spherical cluster 
 containing $N$ mass points
of mass $m_i$ and velocity $v_i$ governed by the above modified Newton equation (\ref{Nvec}). 
%(we shall precise below what is meant here by gravitational equilibrium).}}
According to this equation, the acceleration of an object $i$ interacting with another one of mass $m_j$ is
\begin{equation}
 \frac{dv_i}{dt} \, = \,-\frac{G \, m_j}{r^2_{ij}} + \kappa(t) \, v_i\, \, ,
\end{equation}
\noindent
where $r_{ij}$ is the distance between objects $i$ and $j$    and $\kappa(t) \, = \, 1/t$
  according to Eq.(\ref{kdef}). Multiplying the above equation   by  $v_i= \frac{dr_{ij}}{dt}$, we get
\begin{equation}
 v_i \, dv_i \, = \,-\frac{G \, m_j}{r^2_{ij}} \, dr_{ij}+ \kappa(t) \, v^2_i\, dt \, .
\end{equation} 
\noindent
This equation accounts for only one interaction $i-j$ , and we have to  sum up to account for all
  the gravitational interactions  of the  object $i$ with   the other masses $m_j$ in the cluster. Thus,  we have
\begin{equation}
\frac{1}{2} \,  d (v^2_i ) \, = 
 -   \sum_{j \neq i} \frac {G \,m_j dr_{ij}}{r^2_{ij}}  +\kappa(t) \, v^2_i \, dt \, ,
 \label{imd}
\end{equation}
\noindent
We now integrate the above differential equation.  The system is non-conservative, 
because the additional outwards acceleration term cannot
 be derived as the gradient of  a potential.
The non-Newtonian term is an  ''adiabatic invariant'', since the rate of its effects
is  generally  very slow.  The usual treatment is to consider a limited, but
significant interval of time and to obtain relations between time averages. 
%Thus, we will consider a time average of the last term in Eq.(\ref{imd}) 
%%over  {\bf{a typical crossing time 
 %% $\tau_{\mathrm{cross}} \, = \, 2 \,R /\overline{\mid v\mid}$.}}
The integration of the above equation between time $t_1$ and   time $t_z$, where $z$ is the cluster redshift, gives 
\begin{equation}
\frac{1}{2} \,    \left[ v^2_i(t_z) -  v^2_i(t_1)\right] \, =
   \sum_{j \neq i}  \left[\frac {G \,m_j(t_z)}{r_{ij}(t_z)} - \frac {G \,m_j(t_1)}{r_{ij}(t_1)} \right] + 
  \int^{t_z}_{t_1}  \kappa(t)  \, v^2_i(t) \, dt \, .
 \label{adi}
\end{equation}
\noindent
Let us take $t_1$ as the time of the formation of the system. The effects of the 
 non-conservative term  in the initial  collapse of the system are  limited and we have at equilibrium,
 $\frac{1}{2} \,    v^2_i(t_1) \, =  \sum_{j \neq i}  \frac {G \,m_j(t_1)}{r_{ij}(t_1)} $.  The above expression simplifies
 and summing over all objects $i$, we get
\begin{equation}
\frac{1}{2} \, \sum_{i } \,    v^2_i (t_z) \, =
 \, \frac{1}{2} \, \sum_{i}   \sum_{j \neq i}  \frac {G \,m_j(t_z)}{r_{ij}(t_z)}  + 
 \sum_{i }    \int^{t_z}_{t_1} \kappa(t)  \, v^2_i(t) \, dt \, .
 \label{adiab}
\end{equation}
\noindent
In the above  expression, there is  a factor $1/2$ in front of the double summation
in order not to count twice the same interaction between a mass $m_i$  and the surrounding masses $m_j$.
%The  values of $\tau_{\mathrm{cross}}$  corresponds to some  fraction of the age of the universe (Sect. \ref{Coma}).
We now take the mean over the $N$ masses of the cluster considered to be spherical. The term on the left gives 
$ \frac{1}{2 \, N} \, \sum_{i } \,    v^2_i (t_z) \,=\, (1/2) \, \overline{v^2(t_z)}$,
while the first term on the right in Eq.(\ref{adiab}) leads to   $0.5 \; q' GM_{sc. inv}/R$, 
where $R$ is the cluster radius and  $M_{sc. inv}$
     the mass determined in the scale invariant theory. There $q'$ is an appropriate structural factor,
      which has no influence on the final result, see  Eqs.(\ref{ratio}) or (\ref{ratio2}).
For the non-Newtonian term, we need to know how the velocities are varying with $t$.
 In an  empty space, $v = a \,t$ (Sect. \ref{cons}), while
in a bound two-body system the velocity is  a constant (Sect. \ref{twobody}). Thus, we write  $v(t) = v(t_z) \, (t/t_z)^{\beta}$ with
$\beta$ between 0 and 1. To express the last term in Eq.(\ref{adiab}), we define
\begin{equation}
 F \, \overline{v^2(t_z)}\, \equiv \, \frac{1}{N} \sum_{i }    \int^{t_z}_{t_1} \kappa(t)  \, v^2_i(t) \, dt \, = \,
  %\frac{1}{N} \sum_{i }  \frac{v^2_i(t_z)}{2 \, \beta} \left[1 - \left(\frac{t_1}{t_z}\right)^{2 \beta} \right] \, = \, 
  \frac{\overline{v^2(t_z)}}{2 \, \beta} \left[1 - \left(\frac{t_1}{t_z}\right)^{2 \beta} \right] \, , \; \;  \mathrm{with} \; 
  F  \, = \, \ln\frac{t_z}{t_1} \, ,  \; \mathrm{for} \;\beta=0 \, .
\label{divap}
\end{equation}
\noindent
%The  average velocity dispersion $\overline{v^2(t')}$ is taken  at some appropriate intermediate time $t'$.
% We   express  this velocity dispersion as some fraction $f $   of the observed velocity dispersion 
%$ \overline{v^2(t')} \,=\, f \, \overline{v^2(t_z)}$  with $f < 1$. }}
%  Thus, we write the non-Newtonian term  as
%\begin{equation}
 % \frac{1}{N} \sum_{i }    \int^{t_z}_{t_1} \kappa(t)  \, v^2_i(t) \, dt \, \simeq \, 
 % f \overline{v^2(t_z)} \, \ln\frac{t_z}{t_1}    \, .
 % \label{integr}
 % \end{equation} 
%\noindent
The above mentioned replacements lead to
\begin{equation}
\overline{v^2(t_z)} \, \left( 1 -  2 \, F\right) \, 
 \simeq\, q' \frac{G \, M_{\mathrm{sc. inv}}}{R} \, .
\label{vir}
\end{equation}
\noindent
%$M_{\mathrm{sc. inv}}$ is the total gravitational, dynamical or virial  mass of the cluster in the scale invariant theory}} 
%and $R$ its   radius. The structural factor $q$ depends on the density distribution in the cluster.
%For a uniform density, it would be $3/5$; polytropes of various indices may also be considered.
 %  In fact, $q$   is close to   unity, for example \citet{Abell77} estimated a value $q=0.986$ for the Coma cluster. 
   The observed velocities are radial velocities and  we may write their relations to the total velocities
\begin{equation}
\overline{ v^2_{\mathrm{rad}}} = p \, \overline{ v^2} \, ,  \quad  \mathrm{and}  \quad
 \overline{\mid v_{\mathrm{rad}}\mid} = p' \, \overline{\mid v\mid} \, .
 \label{pp}
 \end{equation}
 \noindent
   For isotropic motions of the galaxies within the cluster,
 we would have  $p=1/3$ and $p'= 1/2$, values which we  adopt below.  
% {\bf{We may use an expression like Eq.(\ref{xi}) for the time $t'$, {\emph{i.e.}} $t' = \xi/H$ and}} 
  We finally write the expression corresponding to the virial theorem in the scale invariant framework
\begin{equation} 
 \overline {v^2_{\mathrm{rad}}} \left( 1  \,- 2 \,  F  \right) \simeq \, p \, q' \frac{G \, M_{\mathrm{sc. inv}}}{R} \, .
\label{virial}
\end{equation}
\noindent    
%There,  the Hubble constant $H$ is taken at age   $t'$.  
%{\bf{The value of $\xi$ at time $t'$ is not the same  as that at time $t_0$
%in  Eq.(\ref{xi}), however it is still of the order of unity.
 %Table 2 of paper I gives the values of ages  and  Hubble parameters $H$ for different redshifts
%for models with   $k=0$ and $\Omega_{\mathrm{m}}=0.30$ and 0.20.
% and thus it permits to have the appropriate $H(z)$ and $\xi$-values.}}
This expression   differs from the classical one by 
the parenthesis on the left side.
The dynamical masses of clusters of galaxies published in literature are based on the standard virial theorem. Some improvements
 in order to take into account the differences of the concentration of galaxies in clusters
 and other differences have been proposed, {\it{e.g.}}  \citet{Rood74}.
 The standard cluster masses   $M_{\mathrm{std}}$ are based  on a relation of the form,
 \begin{equation}
  \overline {v^2_{\mathrm{rad}}}  \simeq  \, p \, q' \, \frac{G \, M_{\mathrm{std}}}{R} \, .
\label{vistand}
\end{equation}
\noindent
The ratio of the standard masses   $M_{\mathrm{std}}$  from Eq.(\ref{vistand}) to the 
masses  $M_{\mathrm{sc. inv}}$ given by the scale invariant theory in Eq.(\ref{virial}) is equal to
\begin{equation}
\frac{M_{\mathrm{std}}}{M_{\mathrm{sc. inv}}} \, \simeq \,
  \frac{1}{1  \,-  \, 2 \,  F } \, .
\label{ratio}
\end{equation}
\noindent
Two protoclusters of galaxies  in a forming stage have been observed at $z=5.7$ \citep{Ouchi05}. 
This  corresponds to  ages between 1.2 and 1.8 Gyr, 
for  models with $\Omega_{\mathrm{m}}$ between 0.3 and 0.1 for $k=0$,
giving an upper limit of $t_0/t_1  \simeq  10$.
 % it is also known that the time of the cluster formation $t_1$ generally represents a significant fraction of the age of the universe. 
%Most clusters have a ratio of
%the Hubble time over their radius crossing time  $R/\overline{\mid v \mid}$ between  about   1  and  10, with a peak at 4 \citep{Jackson75}. 
%For $\beta=0$, the term $2F$ would go from 0 to 4.6, and from for 0 to 0.99 for $\beta=1$.
%Thus, the effect would be negligible only only if the observed layers of the clusters reach equilibrium at the time of the observation.
With $\beta=1$, for $t_z/t_1= 1.5, 2, 4$ and  $10$, we have $\frac{M_{\mathrm{std}}}{M_{\mathrm{sc. inv}}} \, \simeq
2.2, 4, 16$  and $100$ respectively. With $\beta=0$, $\frac{M_{\mathrm{std}}}{M_{\mathrm{sc. inv}}}$ rapidly diverges for $t_z/t_1 > 1.6$.
Thus,  we may conclude that, except for clusters still in formation, 
{\emph{the masses obtained  by the standard virial theorem are often much larger than given by the scale invariant theory}}.
 \\
 
Another estimate of  $F$ can be made by considering an average over an interval of time $\Delta t$ 
equal to   the radius crossing time  $R/\overline{v}$, which often
represents a large fraction of the cluster lifetime, especially when massive clusters are considered.
This offers the  advantage to provide an estimate based on  observed parameters. We write
\begin{equation}
F \, \overline{v^2(t_z)}\, = \, \frac{1}{N} \sum_{i }    \int^{t_z}_{t_1} \kappa(t)  \, v^2_i(t) \, dt \,  \simeq \, 
  \overline{v^2(t')} \frac{R}{ t' \overline{\mid v(t') \mid}}\,
  \simeq  \,f \, \overline{v^2(t_z)} \frac{R}{ t'  f^{1/2} \, \overline{\mid v(t_z) \mid}} \, .
\label{divap}
\end{equation}
The  intermediate time $t'$ is about  $(1/2) t_z$. For $f$, we take 1 as for equilibrium  ($\beta=0$). 
From Eq.(\ref{adiab}), we get
\begin{equation} 
  \overline {v^2(t_z)} \left( 1  \,-  \frac{ 4 \,R}{t_z \,\overline{\mid v(t_z) \mid} } \right)
 \simeq  \, q' \frac{G \, M_{\mathrm{sc. inv}}}{R} \, ,
\label{virial1}
\end{equation}
%\begin{equation} 
 %\overline {v^2_{\mathrm{rad}}(t_z)} \left( 1  \,-  2p' \,  f   \frac{R}{\frac{1}{2} \, t_z \,f^{1/2}\, 
% \overline{ \mid v_{\mathrm{rad}}(t_z) \mid  }}  \right)  \simeq
 % \overline {v^2_{\mathrm{rad}}(t_z)} \left( 1  \,-  \frac{\sqrt{2} \,R}{t_z \,\overline{\mid v_{\mathrm{rad}}(t_z) \mid} } \right)
 %\simeq \, p \, q' \frac{G \, M_{\mathrm{sc. inv}}}{R} \, ,
%\label{virial}
%\end{equation}
\noindent
which leads to the following mass ratio for radial velocities with $p'=1/2$,
\begin{equation}
\frac{M_{\mathrm{std}}}{M_{\mathrm{sc. inv}}} \, \simeq \,
 % \frac{1}{1  \,- 2p' \,  f   \frac{R}{\frac{1}{2} \, t_z \,f^{1/2}\, \overline{ \mid v_{\mathrm{rad}}(t_z) \mid  }}} \, \simeq  
  \, \frac{1}{1  \,-  \,  \frac{2 \,R}{t_z \,\overline{\mid v_{\mathrm{rad}}(t_z) \mid} } } \, .
\label{ratio2}
\end{equation}
\noindent
This confirms that the masses derived in the present theory may be much smaller than the standard masses.
A prediction of the theory is that forming clusters have little or no dark matter.

The ratio  $M_{\mathrm{tot}}/L$  of the mass to the luminosity  of the observed clusters is considered in general. As
the luminosities are essentially independent on
the dynamical state  of the clusters, we also have
\begin{equation}
\left(\frac{M_{\mathrm{tot}}}{L}\right)_{\mathrm{std}} \,  \simeq \, \left(\frac{M_{\mathrm{tot}}}{L}\right)_{\mathrm{sc. inv}} \,
 \frac{1}{1  \,-   \,  \frac{2 \,R}{t_z \,\overline{\mid v_{\mathrm{rad}}(t_z) \mid} } } \,.
\label{ratioml}
\end{equation}
\noindent
The standard mass--luminosity ratios are also  larger than those from the scale invariant framework. 
There are   uncertainties,  nevertheless  these estimates    confirm that 
 some substantial part of the dark matter could   be due to scale invariant effects. 
As the dynamical masses  have contributed
 to ascertain the concept of dark matter, we now examine in two quantitative examples
   what fraction of  the dark matter could possibly  be due  to the above effects.

 \subsection{The case  of the Coma cluster and Abell 2029}  \label{Coma}
 
 Coma  and Abell 2029 are the most studied  massive clusters of galaxies, they both have   about 1000 member galaxies.
  A recent study by \citet{Sohn17} provides  a very complete and detailed study of their luminosity, 
  stellar mass and velocity dispersion functions. For the Coma cluster,
  they found a  mass $M_{200} = 1.29^{+.15}_{-.15} \cdot 10^{15}$  M$_{\odot}$, a radius 
  $R_{200} = 2.23^{+0.08}_{-0.09}$ Mpc (R$_{200}$ and M$_{200}$
 indicate the values up to which  the enclosed density  is equal to
 200 times the critical density). The velocity dispersion $\sigma$ is
  947  ($\pm 31$) km  s$^{-1}$. For Abell 2029, the corresponding data are
 $M_{200} = 0.94^{+.30}_{-.27} \cdot 10^{15}$  M$_{\odot}$, 
  $R_{200} = 1.97^{+0.20}_{-0.21}$ Mpc and  $\sigma=
  973 $ ($\pm 31$) km  s$^{-1}$. 
  We  point out that the radii $R_{200}$ only encompass $\sim 5\%$ of the total volume of these clusters,
  which have observed total radii of about    6 Mpc,  according to the data by \citet{Sohn17}. 
  These authors have published the plot of the 
clustercentric velocities vs.  clustercentric distances for the Coma  and Abell 2020 clusters. These
plots  support that the radial extension of these clusters reaches   6 Mpc.
 %{\bf{It is particularly important to have 
% appropriate cluster radii, since it is known that the mass-luminosity  $M/L$ ratio
 %rapidly increases with the radius considered  \citep{Lewis03}. }}

  For Coma, the redshift is $z=0.0235$  and for Abell 2029 $z=0.0784$. According to the scale invariant cosmological models
  with $k=0$ and $\Omega_{\mathrm{m}} =0.3$ to 0.1  of paper I, this corresponds to ages of about
  13.5 Gyr and 12.8 Gyr respectively, (for these small $z$, different models make  small differencies). 
  As to the radius crossing times, for more  realistic radii of  5 or 6 Mpc,  we get  $=5.16 \; \mathrm{or} \;  6.20$ Gyr (Coma)
   and 5.03 or 6.03 Gyr (Abell 2029) respectively.
  For the factor   $2 \, F = \frac{2 \,R}{t_z \,\overline{\mid v_{\mathrm{rad}}(t_z) \mid} } $, we get 
  $2 \, F  =0.764 \; \mathrm{or} \;  0.919 \,\mathrm{(Coma)} \;    \mathrm{and}  \;  0.786 \; \mathrm{or} 
   \; 0.942 \; \mathrm{(Abell \; 2029)}$. 
The corresponding  estimates of the ratios $\frac{M_{\mathrm{std}}}{M_{\mathrm{sc. inv}}}$ in Eq.(\ref{ratio2}) are
\begin{equation}
\frac{M_{\mathrm{std}}}{M_{\mathrm{sc. inv}}} \, \simeq \, 4.2\;  \; \mathrm{or} \; 
  12.3\;(\mathrm{Coma})  \quad 
  \mathrm{and}  \simeq   \;4.7 \; \mathrm{or} \;  17.2 \; \;(\mathrm{Abell \; 2029})  \, .
\label{rationum}
\end{equation}

\begin{figure*}[t!]
\centering
\includegraphics[width=.60\textwidth]{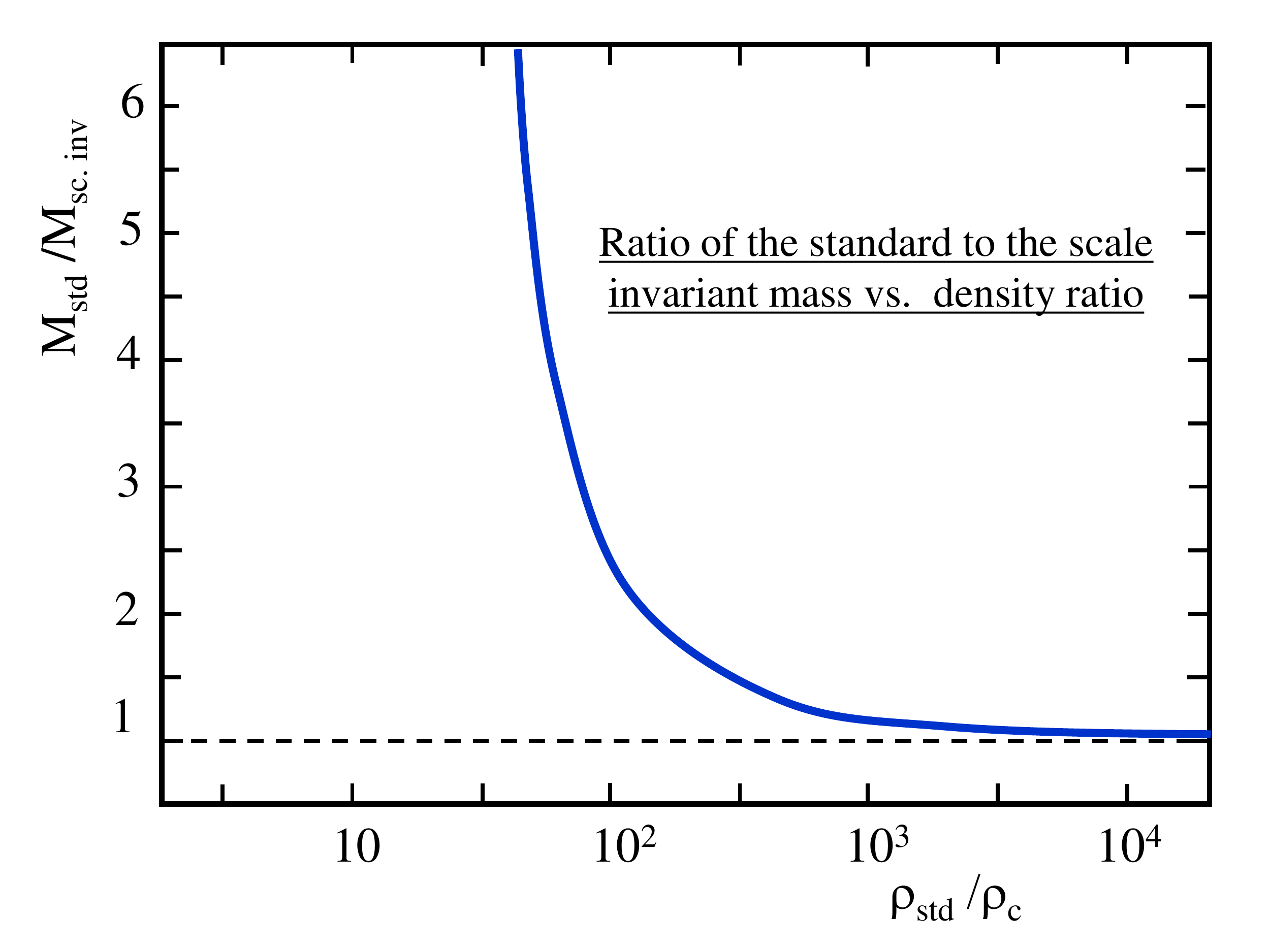}
\caption{The ratio $\frac{M_{\mathrm{std}}}{M_{\mathrm{sc. inv}}}$ as a function
of  the ratio of the standard density to the critical density, both considered 
at  the time  $t_z$ of the cluster redshift $z$. This plot is based on Eq.(\ref{rrr})
with the values of the parameters indicated in the text. It
also applies to the case of no 
or negligible redshifts.}
\label{ratioM}
\end{figure*}

As a matter of facts, the above numerical values likely are not overestimated for two reasons.
1.-- First, the above radii of 6 Mpc may still be too low. For example, in the case of Abell 2029, the concentration of points  
 in Fig. 5 by \citet{Sohn17} may extend up to 8 Mpc. %Such a value of the radii would bring the ratio
 %$\frac{M_{\mathrm{std}}}{M_{\mathrm{sc. inv}}}$ to {\bf{7.6  and 9.3 }}for Coma and Abell 2029 respectively.
 2.-- Secondly, in both clusters  at large clustercentric distances the  velocities are much smaller than in the 
 cluster core. In Fig. 5 and 6 by \citet{Sohn17},  the 
caustics  defining the  velocity  limit decrease by  about a factor of 2 from $R_{200}$ to a distance of 6 Mpc. 
%As at the same time there are less stars in the outer cluster regions,  the resulting decrease of the average velocity is 
%evidently  much smaller than a factor of  2.
 Even if the average velocity is reduced by 5\% or 10\%, this would
significantly increase the ratios of the standard to the scale invariant mass  in both clusters.
 In fact, both effects number 1 and number 2 intervene.

Thus, we see that the dynamical masses estimated in the scale invariant system are smaller
 by a large  factor (of about 4 to  12)
with respect to the standard case. In this context, 
we recall that the baryon fraction from WMAP-7yr and Planck-2013 turns around 0.16 to 0.17.
% which is an upper limit of the the values of 
%$f_{\mathrm{bar}}$ determined  from clusters of galaxies \citep{Gonzalez13,Chiu16}. 
Thus,  with the above numerical  figures, we
see that there would be not much  room, and maybe no room at all, left for dark matter in the context of the scale invariant theory.

We conclude that {\emph{a large fraction of  the dark matter, and possibly the whole of it, is no longer demanded in the framework
of the scale invariant dynamics}}.
% {\bf{We could say that in the scale invariant model, the dynamical mass
%equals the baryonic mass in clusters of galaxies, while in Newtonian dynamics this is a multiple of it.}}
 More detailed 
analyses with extensive numerical simulations of the dynamical evolution of clusters of galaxies 
 in the framework of the scale invariant theory need to be performed  in the future.

\subsection{Mass estimates  in relation with   cluster density}   \label{F}

We now examine  the relation between the excesses of the standard masses (with respect  to the scale invariant results)
and the average cluster densities.
 Expressing  the term  $2F = \frac{4 \,R}{t_z \,\overline{\mid v(t_z) \mid} }$ 
  with $H= \xi/t$  (cf. Eq.\ref{xi}) and the usual
critical density  $\varrho_{\mathrm{c}} \, = \, \frac{3 \, H^2}{8 \pi \,G}$,   we get
\begin{equation}
2F \,= \, \frac{4}{\xi }  \left(\frac{R^2 H^2}{\overline{\mid v(t_z) \mid}^2 } \right)^{1/2} 
\simeq  \, \frac{4}{\xi} \, \left(\frac{2 \,\varrho_{\mathrm{c}}}{ q' \varrho}\right) ^{1/2}\, ,
\end{equation}
\noindent
where we have used  Eq.(\ref{vistand}), also  identifying the quadratic and arithmetic means of the velocities.
 According to  Sect. \ref{vvir}, 
the critical density  $\varrho_{\mathrm{c}}$ must be taken  at the redshift corresponding to $t_z$.
For the mass we take  the standard mass $M_{\mathrm{std}}$, thus the density 
is the corresponding density $\varrho_{\mathrm{std}}$ of the cluster.
The ratio $\frac{M_{\mathrm{std}}}{M_{\mathrm{sc. inv}}} $ may be written

\begin{equation}
\frac{M_{\mathrm{std}}}{M_{\mathrm{sc. inv}}} \,  \simeq \, \frac{1}{1- \frac{4}{\xi} \, 
\left(\frac{\,2 \varrho_{\mathrm{c}}(t_z)}{ q' \,  \varrho_{\mathrm{std}}(t_z)}\right)^{\frac{1}{2}}} \, ,
\label{rrr}
\end{equation}
\noindent
where, as seen above, we  adopt $\xi \sim  1$,  $ q'  \sim 1$. 
We see that the ratio of the standard to the scale invariant 
masses increases for object of  lower densities, consistently with the remarks  in Sect. \ref{magn}. 
For astronomical systems with densities much above the critical density of the universe,
the two mass estimates are similar.  Let us recall that  Table 2 of paper I  allows one 
to estimate $H$  and thus the critical densities at different redshifts . 
%For the Coma and Abell clusters, we got above   ratios $\frac{M_{\mathrm{std}}}{M_{\mathrm{sc. inv}}} \simeq 1.20.

 Fig. \ref{ratioM} shows the  ratio  $\frac{M_{\mathrm{std}}}{M_{\mathrm{sc. inv}}} $ as a function of ratio of
 the cluster density $\varrho_{\mathrm{std}}$ with respect to the critical density at the time  $ t_z$.
We see that the excess  of the standard cluster masses with respect to the values in the scale invariant theory
rapidly diverges for values of the density ratio
 $ \varrho_{\mathrm{std}}/\varrho_{\mathrm{sc. inv}}$  below $10^2$,
  consistently with the results about  the Coma and Abell 2029 clusters.
Finally, we recall that, even within a given cluster, the standard estimates of the (M/L) ratio steeply increase for larger radii \citep{Lewis03}, 
 {\it{i.e.}} for decreasing average internal densities. This remarkable fact  is quite in agreement with the above results 
 and does not demand 
 any peculiar distribution of dark matter according to clustercentric distances.

\section{The rotation curves of galaxies}  \label{sub:rot}

The flat curves of rotation velocities in the external regions of spiral galaxies usually provides another major evidence 
of dark matter, see review by \citet{SRubin01} and further ref. in Sect. 
(\ref{rotation}). The rotation velocities remain about constant instead of decreasing with central distance $r$,  like $ \sim \,1/\sqrt{r}$
 as predicted by the Newtonian law at some distance of an axisymmetric central mass concentration.

\subsection{The two-body problem}  \label{twobody}

We start by  studying  the classical case of the two-body problem in the scale invariant framework,
 following some early developments 
by \citet{MBouvier79}. The specific angular momentum in the classical case is in Cartesian coordinate
 $\bold{x'} \, \times \, \frac{d \bold{x'}}{dt}$, which is constant 
in time. Let us examine  the product  $ \bold{x}  \times \kappa(t) \, \frac{d \bold{x}}{dt}$ and its derivative,
\begin{equation}
\frac{d}{dt} \left( \bold{x}  \times  \kappa(t) \,\frac{d \bold{x}}{dt} \right)\, = 
\, \frac{ d \kappa }{dt}\,\left( \bold{x} \times  \frac{d \bold{x}}{dt}\right)+ 
\kappa(t) \, \frac{d}{dt}\left(\bold{x}   \times  \frac{d \bold{x}}{dt}\right) \, .
\label{derivmom} 
\end{equation}
\noindent 
Now, according to the expression of $\kappa(t)$ in Eq.(\ref{kdef}), one has
\begin{equation}
\frac{d \kappa}{dt} \, = \, - \kappa^2(t) \, .
\end{equation}
Let us develop the two terms on the right of Eq.(\ref{derivmom}),
\begin{equation} 
- \kappa^2(t) \, \left(\bold{x}   \times \frac{d \bold{x}}{dt}  \right) \, = \, -\kappa^2(t) \, \left(x^1 \, \frac{dx^2}{dt}
-   x^2 \frac{dx^1}{dt} \right)+  \mathrm{the \; same \; for } \; (2,3) \;  \mathrm{and} \; (3,1) \\,
\label{n1}
\end{equation}
\begin{equation} 
\kappa(t)\,\, \frac{d}{dt}\left(\bold{x}  \times  \frac{d \bold{x}}{dt}\right) \, = \,\kappa(t) \, \left( \frac{dx^1}{dt}\frac{dx^2}{dt}+ x^1 \frac{d^2 x^2}{dt^2}
-   \frac{dx^2}{dt}\frac{dx^1}{dt}-  x^2 \frac{d^2  x^1}{dt^2} \right)+  \mathrm{the \; same \; for } \; (2,3) \;  \mathrm{and} \; (3,1) \, .
\label{n2}
\end{equation}
\noindent
In this last expression, the first and third terms on the right cancel each other, the second and fourth are according to Eq.(\ref{cartes}),
\begin{equation}
x^1 \, \frac {d^2  x^{2}}{dt^2} \, = \, - \frac{G \, M}{r^2} \, \frac{x^{1} \, x^2}{r}   + \, \kappa(t)  \, x^1 \frac{dx^{2}}{dt} \, .
 \label{cartes2}
 \end{equation}
\begin{equation}
x^2 \, \frac {d^2  x^{1}}{dt^2} \, = \, - \frac{G \, M}{r^2} \, \frac{x^{1} \, x^2}{r}   + \, \kappa(t)  \, x^2 \frac{dx^{1}}{dt} \, .
 \label{cartes1}
 \end{equation}
 \noindent
 Now, we can write the complete expression
 \begin{eqnarray}
\frac{d}{dt} \left(\, \bold{x} \, \times \, \kappa(t) \, \frac{d \bold{x}}{dt} \right)\, = \, -\kappa^2(t) \, \left( x^1 \, \frac{dx^2}{dt}
-   x^2 \frac{dx^1}{dt} \right) +
\kappa(t) \left(- \frac{G \, M}{r^2} \, \frac{x^{1} \, x^2}{r}+ \kappa(t)\, x^1 \frac{dx^{2}}{dt}\right)  \\  \nonumber
-\kappa(t) \,\left( - \frac{G \, M}{r^2} \, \frac{x^{1} \, x^2}{r}+  \kappa(t)  \, x^2 \frac{dx^{1}}{dt} \right) 
+  \mathrm{the \; same \; for } \; (2,3) \;  \mathrm{and} \; (3,1) \, = \, \bold{0} \, .
\label{mom}
\end{eqnarray}
\noindent
The two   Newtonian terms cancel each other and the same for the other terms.
Thus, the  above equation expresses the angular momentum conservation in the scale invariant framework.
The vector $\left(\kappa(t)  \, \bold{x}  \times \frac{d \bold{x}}{dt} \right)\, $ is always 
orthogonal to the orbital motion, which indicates that the problem is 2--dimensional. 
The angular momentum conservation  writes in polar coordinates $(r, \vartheta)$,
\begin{equation}
\kappa(t) \,  r^2 \, \dot {\vartheta} \, = \, L  \, = \, \mathrm{const.}
\label{ang}
\end{equation}
\noindent
It is a scale invariant term.
 At a fixed time, the above expression  is similar to the
usual conservation law. Now, the equation of motion (\ref{Nvec}) writes in the two polar coordinates
 \begin{equation}
 \ddot{r} - r \, \dot{\vartheta}^2 \, = \,- \frac{G \, M}{r^2} +\kappa(t) \, \dot{r} \, ,
 \label{Nr}
 \end{equation}
 \begin{equation}
 r \, \ddot{\vartheta} + 2 \, \dot{r} \, \dot{\vartheta} \, = \, \kappa(t) \, r \, \dot{\vartheta}
 \label{Ntheta}
 \end{equation}
 \noindent
 The mass $M$ is the mass in the scale invariant framework, see expression (\ref{mass}).
 We easily verify the compatibility of expression (\ref{ang}) with the above equation (\ref{Ntheta}). The radial equation (\ref{Nr})
 can be expressed with $L$,
 \begin{equation}
 \ddot{r} -   \left(\frac{L}{\kappa(t)}\right)^2 \,\frac{1}{r^3} + \frac{G \, M}{r^2} - \kappa(t)  \, \dot{r} \, = \, 0 \, .
 \label{NL}
 \end{equation} 
 \noindent
 We may transform the time derivatives  into derivatives with respect to $\vartheta$, with
 $\frac{dr}{d\vartheta} = \frac{\dot{r}}{\dot{\vartheta}}= \frac{\dot{r}}{L} \kappa(t) \, r^2$ we have,
 \begin{equation}
 \dot{r} \, = \, \frac{L (dr/d\vartheta)}{\kappa(t)  \, r^2}\, \quad \mathrm{and} \quad 
 \ddot{r} \, = \, \frac{L^2}{H^2\, r^4}\left(\frac{d^2 r}{d\vartheta^2}- 2 \frac{(dr/d\vartheta)^2}{r}   \right)
 -\frac{\dot{\kappa(t) } \, L \, (dr/d\vartheta)}{\kappa^2(t) \, r^2} \, 
\label{rr} 
 \end{equation}
 \noindent
 These replacements lead to an equation in $(r, \vartheta)$  giving the curve described by a test particle in the central field of the scale invariant mass. Most remarkably, the last term in the modified Newton equation (\ref{NL}) simplifies with the last one in  expression  (\ref{rr})
 for $\ddot{r}$, and we have
 \begin{equation}
 \frac{L^2}{\kappa^2(t)  \, r^4} \left(  \frac{d^2 r }{d \vartheta^2}- \frac{2}{r} (\frac{dr}{d\vartheta})^2 \right)
 - \left(\frac{L}{\kappa(t) } \right)^2 \frac{1}{r^3} + \frac{GM}{r^2}  \, =  \, 0 \, .
 \label{Binet1}
 \end{equation}
 \noindent
 This allows us with the transformation  $\rho \, = 1/r $ to write
 \begin{equation}
 \frac{d^2 \rho}{d\vartheta^2} + \rho \, = \, \frac{G \, M \, \kappa^2(t)}{L^2} \,.
 \label{Binet2}
 \end{equation}
 \noindent
 This expression is identical to the classical Binet equation except for the $\kappa$--term on the right. Thus, we may 
 immediately write the solution $\rho = (1/r_0) + C \, \cos(\vartheta)$ or
 \begin{equation}
 r  \, = \, \frac{r_0}{1 + e \cos(\vartheta)}, \quad
 \mathrm{with} \quad r_0 \, = \, \frac{L^2}{G \, M \, \kappa^2(t)} \, .
 \label{sol}
 \end{equation}
\noindent
There, $r_0$ is the radius of a circular orbit (for $e=0$). It is not a scale invariant quantity. Recalling 
once more that the Einsteinian mass is $M'= \lambda \, M$, we see that $r_0$ grows like $t$, consistently with the
basic relation (\ref{lambda}).
The eccentricity $e$ is given by
\begin{equation}
e \, = \,  C \, \, \frac{L^2}{G \, M \, \kappa^2(t)} \, , \quad  i.e.  \quad  C= e/r_0 \, .
\label{e}
\end{equation}
\noindent 
%$C$ depends on the location of the periastron. 
We verify that the eccentricity  $e$ is scale invariant, which is satisfactory.
The above equation (\ref{sol}) is that of a conic, ellipse, parabola or hyperbola depending on the eccentricity, however
with a secular variation of the orbital radius $r_0$, or semi-major axis as shown below.

 The solutions of the two-body problem are similar to those of the standard case, with in addition a slow secular variations of the orbital 
 radius. More generally, if we consider the semi--major axis $a$ of an orbital motion,
 \begin{equation}
 a \, = \, \frac{r_0}{1-e^2} \, ,
 \label{a}
 \end{equation}
 \noindent
 we have from Eqs.(\ref{sol}) and (\ref{a}), together with Eqs.(\ref{LE}) and (\ref{kdef}), 
 \begin{equation}
 \frac{\dot{a}}{a} \, = \, \frac{\dot{\lambda}}{\lambda}- 2 \frac{\dot{\kappa}}{\kappa}\,= 
 - \frac{\dot{\kappa}}{\kappa}\, = \, \frac{1}{t} \, .
 \label{aa}
 \end{equation}
 \noindent
 Thus, we see that the semi-major axis increases linearly with time $t$.  
The behavior of the circular velocity $v_{\mathrm{circ}}$ is also interesting.
 From   Eq.(\ref{ang}) of the conservation of the angular momentum,
we get 
\begin{equation}
v_{\mathrm{circ}} \, = \, r_0 \, \dot{\vartheta} \, = \, \frac{L}{\kappa(t) \, r_0} \, = \, const.
\label{velocity}
\end{equation}
\noindent
The constancy results from the fact that $\kappa(t)$ behaves like $1/t$ and $r_0$ like $t$, $L$ being  a constant.
This is consistent with the fact that the gravitational potential is an invariant as shown by Eq.(\ref{Phi}).
The constancy of the circular velocity over the times is of great importance for the study of the rotation curves of galaxies below.
From the conservation  law (\ref{ang}), we also see that the orbital period $P$  similarly varies like $\dot{P}/P \, = \, 1/t$. 
This is also evident since 
the radius increases linearly and both the eccentricity and the circular velocity are constant.

Thus, the scale invariant two-body problem leads essentially to the same solutions as the Newtonian case, with a slight supplementary 
outwards expansion at a rate which is not far from the Hubble expansion. These conclusions consistently come from the  hypothesis 
we have made (see Sect. 1). Now, whether this corresponds to Nature or not, can only be decided on the basis of
careful comparisons with observations.

\subsection{An application to the outer  rotation  curves of galaxies}   \label{rotation}

The rotation curves of nearby spiral galaxies, {\it{i.e.}} the circular velocities as a function of the galactocentric distances $r$, generally  remain 
flat in the outer regions, instead of having a Keplerian decrease like   $ \sim \; 1/\sqrt{r}$, as expected if
    most of  the mass lies in inner regions.
   The velocity determinations are
 mainly based on optical observations of H$\alpha$, NII and SII lines and on radio observations of HI and CO lines. 
 There is a long history of the  problem of the flat rotation curves, as reviewed by \citet{SRubin01}, who report
that already in 1940, Oort noticed ''... the distribution of mass [in NGC  3115]
appears to bear no relation to that of the light.''
Such facts were further  confirmed by  other precursors. From  a sample of 10 high--luminosity 
spiral galaxies, \citet{Rubin78} stated that ''all rotation curves are approximately flat, to a distance as great as $r= 50$ kpc.''
The sample was  extended to 21 galaxies \citep{Rubin80},
further supporting the previous conclusions. Nowadays, the observations of thousands of galaxies
confirm the difference of the matter and luminosity distributions and  support the existence of a halo of dark matter around
  the Milky Way and other galaxies, {\it{e.g.}}  \citet{Persic96,SRubin01,Sofue12,Huang16}.

We concentrate on the case of the Milky Way where the rotation curve is known to the largest distances from the center.
On the basis of the velocities of  about 16 000 red clump giants in the outer disk, as well as $\sim$ 5700 halo K giants in the 
halo,  \citet{Huang16} have constructed the rotation  curve of the Milky Way up to about 100 kpc. The  average data as a function of the galactocentric distance are given in their Table 3, which indicates the various segments of the curve and the source of their measurements.

\begin{figure*}[t!]
\centering
\includegraphics[width=.99\textwidth]{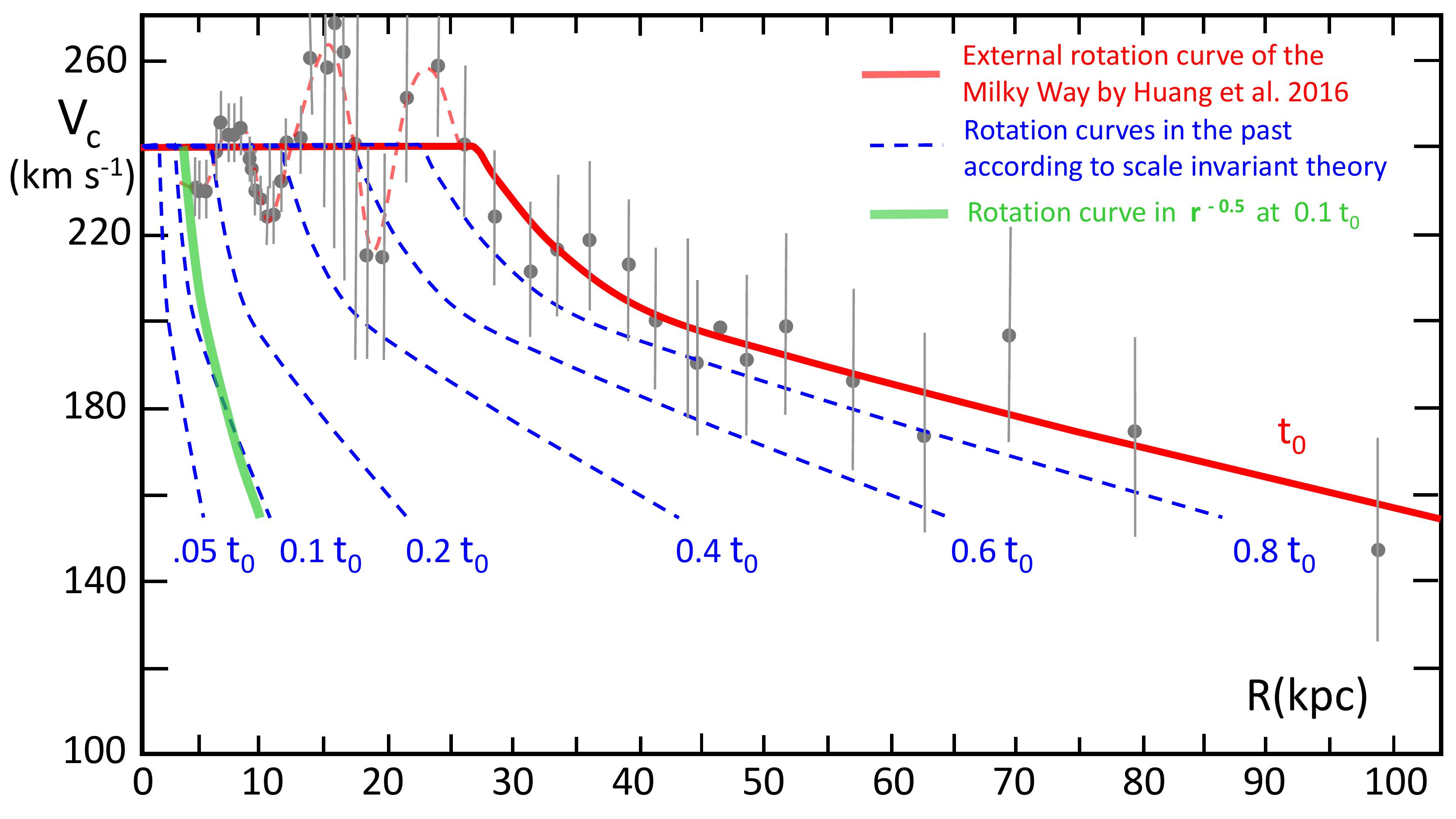}
\caption{Evolution of the rotation curve of the Milky Way. The gray points are the observed velocity
 averages by \citet{Huang16} with their error bars and the thick
red line represents the corresponding average rotation curve. The thin broken red line describes the undulations in the globally flat 
part of the distribution.
The blue broken lines show
 the  rotation curves  predicted by   the scale invariant theory for  different past epochs expressed 
  in fraction of the present age  $t_0$.  The thick green line shows a Keplerian curve  in  $1/\sqrt{r}$  at a time corresponding to
  10\% of the present age. We see that this Keplerian  curve is close to a distribution consistent with the scale invariant 
  theory in an early epoch.}
\label{MW}
\end{figure*}

 The curve by \citet{Huang16}  is illustrated in Fig. \ref{MW}, it shows a flat 
rotation  curve with a circular velocity of 240 km s$^{-1}$ up to galactocentric distances $R$ of about 25 to 30 kpc 
and then it slowly decreases down to 150 km s$^{-1}$ at 100 kpc.
 There is also some prominent dips at  $R=11$ and 19 kpc,  
(represented in the  light broken red line in Fig. \ref{MW}). The error bars on the velocities  are rather small
 ($\sigma \approx 7$ km s$^{-1}$) for $R$ between 4.6 kpc and about 13 kpc, so that the dip 
at 11 kpc appears as very significant. From $R=15$  to 20 kpc, the error bars are much larger so that the dip at 19 kpc
may be less significant. Nevertheless, in view of the small amplitude of the dips with respect to the velocity, 
 the rotation curve may be considered as  globally flat up to at least 25 kpc \citep{Huang16}. We note that
 the 11 kpc dip is often  interpreted as due to a ring of dark matter at that location. 
The dip at 19 kpc may have the same origin,
however it could also be artificial  due to the use of different data sets. We note that \citet{Rubin78} already pointed outed
secondary velocity undulations  in various rotation curves, with   rotational velocities lower
 by about 20 km s$^{-1}$ on the inner edge than on the outer edges of spiral features.
 
 The decrease in the external regions  reaches about 100 km s$^{-1}$,
  it is about five times larger than the error bars. Moreover, it 
 is supported by all measurements beyond about a galactocentric distance $R \approx 25$ kpc.
 The observed points then  form a rather smoothly  decreasing curve.

The red  curve in Fig. \ref{MW} is the velocity distribution at the 
present cosmic time $t_0$. (In the cosmological models of paper I, the present age is fixed to $t_0=1$ which corresponds 
to 13.8 Gyr. The correspondence between $t_0$ and $H_0$ is expressed by Eq.(23) with the
appropriate $\xi$-values). We can find the corresponding velocity distributions 
at past epochs, 0.8 $t_0$, 0.6 $t_0$, 0.4 $t_0$, etc... by applying the properties of Eqs.(\ref{aa}) and  (\ref{velocity})  
derived from the equivalent  Binet equation (Eq.\ref{Binet2}) in the scale invariant framework.
 At past epochs, the radii were smaller, while the circular velocities kept constant. Thus,
we apply these simple evolution laws to the present rotation curve to deduce the curves at past epochs.
Of course, this does not preclude the  various dynamical effects which currently are at work in galaxies to be simultaneously operating:
interactions due to spiral waves, effects of bars, non-axisymmetric perturbations, radial motions, cloud collisions, mergers, etc.  
For now, we ignore these various effects in order
 to just examine the consequences  of scale invariance.
  %Depending on the nature of the results,  complex numerical models
%of the dynamical evolution of galaxies accounting for the various above effects together with those of scale invariance should be developed.

Fig. \ref{MW} shows that  at earlier epochs the outer velocity distributions derived from the scale invariant  predictions 
were increasingly steeper with decreasing time.
 At the same time, the Galaxy was more compact.
 % Interestingly enough, in the early times, the resulting velocity curves  become
% close to a Keplerian law. As an example, this is  shown  for time $0.1 \; t_0$ in green in Fig. \ref{MW}. 
The galaxy formation occurred on a  relatively short  timescale  compared to the age of the Universe. At  a given location,
the infalling matter stops
its collapse  when the centrifugal force equilibrates gravity, thus establishing a Keplerian law. Later, during the aging of the Galaxy,
the  dynamical effects of scale invariance  intervene, leading to a flatter distribution. %They are accompanied by
 %the classical dynamical process, which may somehow influence the velocity distribution.
 
  Let us consider that the initial Keplerian velocity distribution was of the form  $v(r, t_{\mathrm{in}}) \, =
 \,  v_{\mathrm{in}}\, \sqrt{r_{\mathrm{core}}(t_{\mathrm{in}})/r(t_{\mathrm{in}})}$, with the assumption of a relatively constant
 circular velocity $v_{\mathrm{in}}$ up to a distance $r_{\mathrm{core}}(t_{\mathrm{in}})$, 
 followed beyond $r_{\mathrm{core}}$ by a Keplerian decrease.  As time is going, the orbital radii increase
 by a factor $\lambda(t)$, thus at time $t$ the velocity distribution becomes,
 \begin{equation}
  v(r, t) \, = \, v_ {\mathrm{in}}\, \sqrt{\frac{\lambda(t) \, r_{\mathrm{core}}(t_{\mathrm{in}})}{\lambda(t) \, r(t_{\mathrm{in}})}} \, =
  \, v_ {\mathrm{in}}\, \sqrt{\frac{ r_{\mathrm{core}}(t)}{\, r(t)}} \,
   \, , \quad \mathrm{for}  \quad
   r(t) \, > \,  r_{\mathrm{core}}(t) \, .
  \label{vsqrt}
  \end{equation} 
  \noindent
  We see that the scale transformation conserves the Keplerian law in $1/ \sqrt{r}$. As a matter of fact, 
  the velocity distribution found by \citet{Huang16} in the external regions of the Galaxy is  close to a
  Keplerian law starting from  $ R \approx 30 $ kpc. Consequently,   
  the curve at  past epochs, like  $0.1 \, t_0$, derived
   by a backwards scaling  from the present curve by \citet{Huang16}, is also close to a steep 
   Keplerian distribution as shown in Fig. \ref{MW}.

 Two important remarks need to be done. A) There is a variety of the rotation curves of galaxies as shown by \citet{SRubin01}. 
 The available data  generally concern radial extensions
 smaller than 20 or 30 kpc. Two very massive galaxies, UGC 2953 and UGC 2487, have 
  been observed up to radial distances of 60 and 80 kpc respectively \citep{Sanders07,Famaey12}. 
  Over these ranges, they only show a decline 
 of 40--50 km s$^{-1}$, smaller than the decrease of about 100 km s$^{-1}$ for
 the Milky Way.  However, 
 these two galaxies are among the most massive and  fastest rotating galaxies, with maximum velocities of about 
  300 and 380 km s$^{-1}$ respectively,
 much higher than in the Milky Way or in the galaxies studied by \citet{SRubin01}. Thus, it would be extremely interesting
 to know the rotation curves in the further outer layers of these extreme galaxies to see whether the decrease goes on, 
 and whether their data can be interpreted in the context of the scale invariant dynamics. 
 B) We also note that a remarkable correlation between the radial acceleration  derived from the rotation curves  
 and the distribution of baryons has been found \citep{McCaugh16,Lelli17},
 implying that the dark matter is fully specified by the baryons.
 The obtained relation indicates the absence of dark matter at high
 acceleration and a systematic deviation for acceleration lower than
 about $10^{-10}$ m s$^{-2}$. These findings that  imply deviations from
 standard dynamics at the  lower densities might provide further tests 
 of the scale invariant dynamics and will be studied in a further work,
 (I am very  indebted to the referee for these remarks).

Thus, we tend to conclude that the relatively flat rotation curves of spiral galaxies is an age effect from
the mechanical laws, which account for the scale invariant properties of the empty space at large scales.
These laws predict that the circular velocities remain the same, while a very low expansion at a rate not far from the Hubble 
rate progressively extends the outer layers, increasing the radius of the Galaxy and  decreasing its surface density
 like  $1/t$ when account is given to Eq.(\ref{mass}).
 It is interesting  that both the mass excesses  derived from the virial in clusters of galaxies and 
from the flat  rotation curves tend to  find an explanation within the scale invariant theory.
 In both cases, there is apparently no need of dark matter and unknown particles.

\begin{figure*}[t!]
\centering
\includegraphics[width=.60\textwidth]{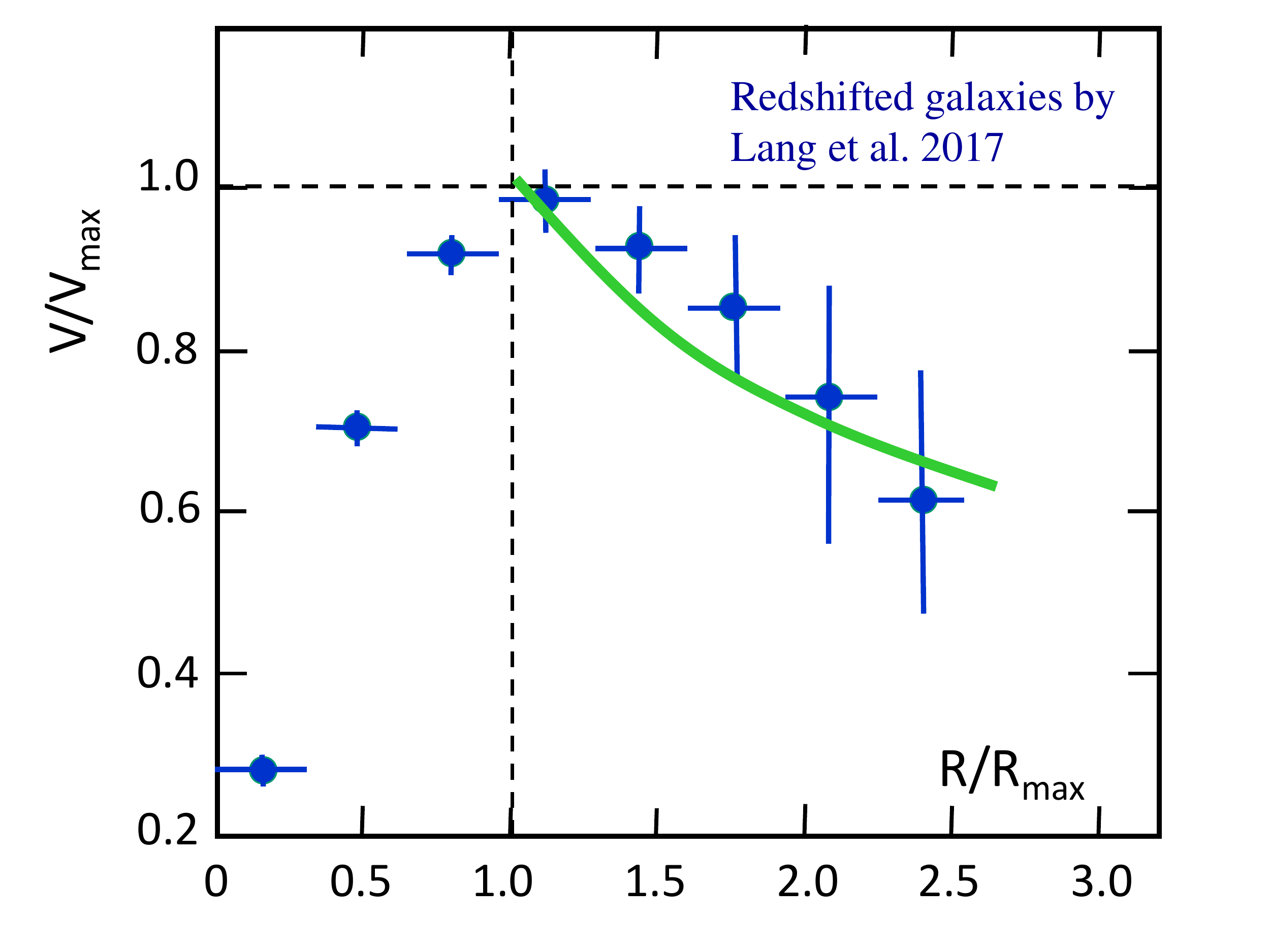}
\caption{Stacked rotation curves with error bars from \citet{Lang17} showing the normalized  velocities vs. the normalized radii
({\it{i.e.}} the radii with respect to   $R_{\mathrm{max}}$,
 the radius where the maximum velocity is reached). The thick  green line shows  the Keplerian curve starting at the maximum 
velocity.}
\label{Lang}
\end{figure*}

\subsection{The age effect derived from \citet{Genzel17} and \citet{Lang17}}

The age effect in the rotation curves of galaxies is  nicely confirmed by recent works by \citet{Genzel17} and \citet{Lang17}.
 % and \citet{Wuyts16}
  Six star forming galaxies in the range $z=0.8 - 2.4$ were studied in details by \citet{Genzel17}, and a sample of 101 other galaxies 
  between $z=0.6$ and 2.6 by \citet{Lang17}. 
The rotation curves they derive  for these early objects show, with a high statistical significance, 
 that the rotation velocities are not constant, but decrease in a compelling way with radius. 
 They show that no dark matter is required to interpret the data, the rotation
 curve is consistent with  a pure baryonic disk. Even at 
 a distance of several  effective radii, the authors 
  find that the dark matter fractions are modest or negligible, the results
  being essentially insensitive to the $M/L$ ratios. 
  
  Fig. \ref{Lang} shows the stacked rotation curves  with their error bars as derived by \citet{Lang17}. 
  The points outwards the radius with maximum velocity show a decrease, which is not far from a $1/\sqrt{r}$ Keplerian curve (in green).
  Most of the galaxies of the sample are observed at epochs before the peak of star formation. 
  This  shows that the usual  flatness of the rotation curve is a characteristic of  the present epoch, but is a property
   absent in the early stages.
  We emphasize that it is a bit worrying that the concentrations of dark matter,  
  in the potential wells of which galaxies are supposed to form,
  are not present in epochs close to the formation time.
  Moreover, there is    a progression in the  presence of dark matter in spiral galaxies with time,
  since the observations by \citet{Wuyts16} indicate that galaxies at $z=1$ contain more dark matter
   than galaxies at $z=2$, and in turn the local present galaxies show more dark matter than those at $z=1$.
 
  Above in Sect. \ref{rotation}, we have described a possible sequence in the dynamical evolution: 
  cloud collapse -- equilibrium  -- steep Keplerian velocity distribution --
   secular evolution according   to  Eqs.(\ref{aa}) and (\ref{velocity}) -- flatter rotation curve of galaxies.  This scenario
   appears to account simultaneously 
   for the observations of \citet{Genzel17} and \citet{Lang17}, concerning the steep Keplerian rotation curve and the
   absence of dark matter  at significant redshifts $z \geq 2$, for the intermediate situation at medium redshifts $z \approx 1$  
   \citep{Wuyts16}, as well for the present flat rotation curves of most 
   galaxies.  These results appear to give some support to the above scale invariant dynamics based on 
   the modified Newton equation (\ref{Nvec}).

   Now, we may wonder whether the progressive flattening of the galaxy rotation curve is the only consequence of the scale invariant 
   stellar dynamics. As a matter of fact, the velocity dispersion, in particular in the so-called ''vertical direction'' shows   
   a strong increase with the age, the age-velocity dispersion relation (AVR) \citep{Seabroke07}.
   This relation has received
   a  variety of explanations over the last decades without any clear consensus, see for example \citet{Kroupa02} and \citet{Kumamoto17}.
    The velocity dispersion in the galactic plane is dominated by the effects of 
   spiral waves as well as by the collisions with giant molecular clouds which are strongly concentrated in the galactic plane. 
  However,  in the directions orthogonal to the plane, there is little interaction since the stars spend most of their lifetimes  out 
   of the galactic plane  \citep{Seabroke07}.  Thus we may wonder whether  the  secular effects of  scale invariance 
    may  play some role. The answer is positive, this  problem is examined in the  Appendix below. 
    
    We also emphasize that the two problems of velocity dispersion and rotation curves are related. 
    The vertical dispersion is an expression of the  support
    in the vertical direction, while the rotation curves express the mechanical support in the horizontal direction.
    The results by \citet{Genzel17} and \citet{Lang17} show that the horizontal support is increasing  with age, and the 
    AVR shows a similar result for the vertical support.  Thus, both mechanical supports of the Galaxy, 
    vertical and horizontal, show an increasing trend with age.

   \section{Conclusions and perspectives}

\begin{figure*}[t!]
\centering
\includegraphics[width=.70\textwidth]{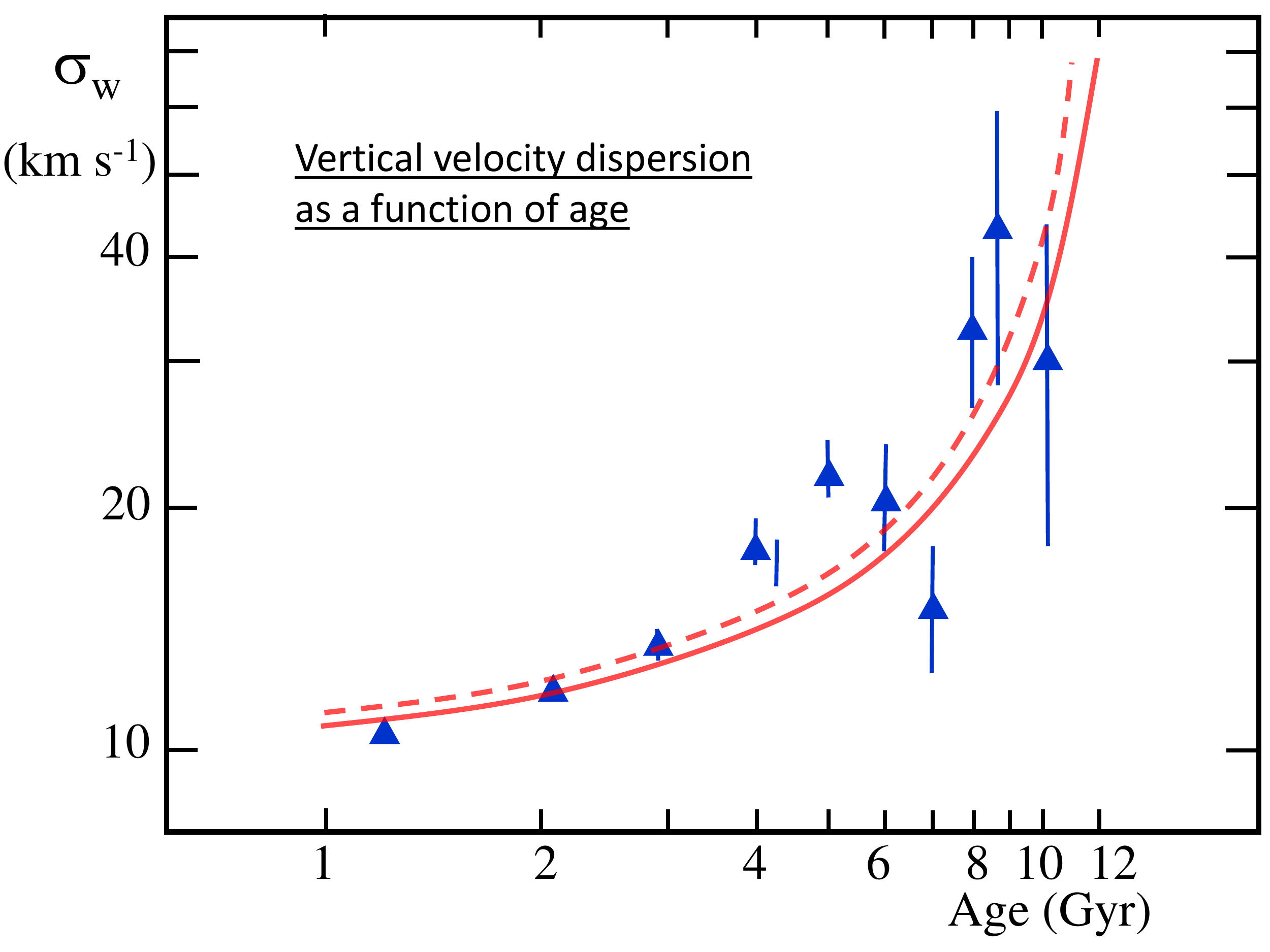}
\caption{The vertical dispersion $\sigma _W$ as a function of the age of the stellar populations.  The blue triangles with 
the error bars result from the analysis by  \citet{Seabroke07} of the observations by \citet{Nordstrom04}. 
The continuous red curve shows the predictions of the scale invariant dynamics according to relation  (\ref{W}) for an age 
  of the universe of 13.8 Gyr. The broken red curve accounts for the fact that the Galaxy formed about 400 Myr after the 
  Big Bang \citep{Naoz06}.
A  vertical dispersion of  10 km s$^{-1}$ is assumed at the present time. }
\label{dispersion}
\end{figure*}

   There is progressively an accumulation of tests supporting the hypothesis of the scale invariance of the empty space at large scales,
   see also \citet{Milgrom09}.
   Firstly, there are the various cosmological tests \citep{Maeder17a} mentioned in the introduction,
   as well as the  test on the past CMB temperatures vs. redshifts \citep{Maeder17b}.
Now, the  studies of the clusters of galaxies, of  the rotation curves of the Milky Way and of high redshift galaxies, 
as well as  of the vertical velocity dispersion of stars in the Milky Way,   all appear   positive.
 The long standing problems of the dark energy \citep{Maeder17a} and
 of the dark matter  may possibly find some solutions in terms of scale invariance.
 In this context, it is noteworthy that it has been claimed that halos of dark matter particles are inconsistent with 
 a large variety of astronomical observations and in particular given the absence in the data
 of evidence for dynamical friction on the motions of galaxies  due to these particles \citep{Kroupa15}.
 
  These  various results are    encouraging and the hope is that they will stimulate future works.
  The list of problems that   await further  studies is  long.
 In this context, we again stress a central point of methodology, 
the tests to be valuable need to be internally coherent
and not use ''observations'' implicitly  involving in their derivations  other cosmological models or mechanical laws.\\

Acknowledgments: I want to express my best thanks to D. Gachet, R. Mardling,  G. Meynet and S. Udry for their 
support and  encouragements.

\appendix

   \section{The  vertical dispersion of stellar velocities in the Galaxy}

   We  examine here the so--called problem of the age--velocity dispersion relation (AVR). This problem is in general not 
    considered as an indication of dark matter, however  we shall see that it may provide  another possible valuable
   indication about the effectiveness of scale invariant dynamics.
   Three velocity components of
   the stellar velocities in the Galaxy are usually defined in stellar dynamics: 
   component $U$ towards the center, $V$ in the direction of the galactic rotation,
   $W$ orthogonal to the galactic plane. The AVR problem is that of explaining why the velocity dispersion, in particular for
   the $W$--component, considerably  increases with the age of the  stars considered, see for example   \citet{Seabroke07}.  
   Continuous  processes, such as spiral waves, collisions with giant molecular clouds, etc... are active in
   the disk plane and may effectively influence the  stellar velocity distributions.
   However as emphasized by these authors, vertical heating
   (the increase of the dispersion  $\sigma_{\mathrm{W}}$) is unexpected, since   stars spend most of their lifetime
    out of the galactic plane. Thus, in order they continuously receive some heating during their lifetime,
    there should be  some heating process  also active away from the galactic plane.
   
   The problem of the AVR and of the  vertical heating   already has a long history. A
   relation was first discovered   by \citet{Stromberg46}, in terms of a relation between velocities  and stellar masses. 
   It was further analysed  by \citet{Spitzer51} who studied the growth of the dispersion due to stellar collisions with giant molecular clouds, 
   an effect also advocated by  several followers.
   \citet{Seabroke07} performed a careful analysis of the extensive data set  by \citet{Nordstrom04} and examined
   the time behavior of the various heating processes.  Their data points are given as blue triangles in 
   Fig. \ref{dispersion} showing  the vertical dispersions as a  function of age. 
   Seabroke and Gilmore  pointed out that the heating by giant molecular clouds should saturate after some time
   and that the dispersion would no longer increase. They give evidence that the vertical heating is continuous  throughout the 
   galaxy lifetime.  Interestingly enough, they noticed the possible effect of a merger about 8 Gyr ago, visible as an outlier point 
   in their  figures (see also Fig. \ref{dispersion}).
   Among the other mechanisms considered, 
 we may mention   the heating by the gravitational field of  spiral waves
 \citep{Barbanis67,Simone04}, the
  heating by an unknown diffusive process \citep{Wielen77}, by massive halo black holes 
 \citep{Lacey85},  by  mergers  of dwarf galaxies \citep{Toth92}, 
 by the effects of evaporating star clusters \citep{Kroupa02}  when the star clusters which form
 expel their residual gas causing the born stars to expand from the embedded cluster,
  also   the  effects of the  evolution of the interstellar  medium
 in the Galaxy  has been advocated by \citet{Kumamoto17}. % Calculations by ... support the view 
 %that the heating by transient spiral waves is the dominant process in the plane near the Sun, but it may progressively saturate.
   As stated by these last authors, there is no consensus on the primary source of the AVR.
   
   Let us study the effects of the scale invariance on the ''vertical'' velocity dispersion perpendicular to 
   the galactic plane following  \citet{Magnenat78}.
   One may assume relatively small oscillations and  far enough from the galactic center. 
   Thus,  the potential perpendicular   to the galactic plane is separable and  the vertical force law 
   $K_z$ is linear in $z$. Taking  the acceleration term as in  Eq.(\ref{cartes}) into account,
   the equation of motion for the $z$--component becomes 
   \begin{equation}
   \ddot{z} \, - \, \frac{1}{t} \, \dot{z} + \omega^2(t) \, z = 0 \, ,
   \label{equz}
   \end{equation}
   \begin{equation}
   \mathrm{with} \quad  \omega^2(t) = \left(\frac{\partial K_z}{\partial z}\right) \, = 4 \, \pi \, G \varrho \, .
   \label{oo}
   \end{equation}
   \noindent
  There, $\varrho$ is the matter density in  g$\cdot$ cm$^{-3}$ at the level of the galactic plane.
   Care has to be given that $\omega^2(t)$ behaves like $1/(t^2)$ according to relation (\ref{mass}) and the    preceding remarks.
   Thus, the oscillation periods increase linearly with time. % see also Eq. (\ref{aa}).
   The analytical solutions of (\ref{equz}) for  $z$ and $\dot{z}$ are,
   \begin{equation}
   z \, = \, \frac{z_{\mathrm{in}}}{t_{\mathrm{in}}} \, t \, \sin (s \, \ln t+\varphi)\, , 
   \quad \mathrm{with} \quad  s\, = \, \sqrt{\omega^2_0 \, t^2_0-1} \, .
   \label{z}
   \end{equation}
   \begin{equation}
  W\, \equiv \,  \dot{z} \, = \,  \frac{z_{\mathrm{in}}}{t_{\mathrm{in}}} \, \left[\,\,  \sin (s \, \ln t +\varphi) + 
    s \cos (s \, \ln t+\varphi) \,\right] \, .
\label{w}
\end{equation}
\noindent
There, the initial and present  values have  indices  ''in'' and ''0'' respectively, $s$ is a number depending on the relative difference 
between the present age and the oscillation period.  Eq.(\ref{z}) shows that 
 the maximum amplitude $z_{\mathrm{max}} \, = \,(z_{\mathrm{in}}/t_{\mathrm{in}}) \, t \, $ 
reached by a given star  increases with  the cosmic time. %  we remark that this is similar with the behavior predicted by (\ref{aa}). 
The velocity of a star born at a given time  always keeps the same velocity  
$W\,  = \,  \frac{z_{\mathrm{in}}}{t_{\mathrm{in}}} \, s \, $  when crossing
orthogonally the galactic plane. %  (see also  Eq. (\ref{velocity}). 
As a matter of fact, this (surprising) behavior of the 
velocity  is consistent with the fact that both $z_{\mathrm{max}}$  and the period of oscillation increase linearly with time.
However, the constancy of the velocity of a given star does not mean that stars born at different times
in the past (even if  born at the same $z_{\mathrm{in}}$) will have the same velocity at present time $t_0$! 
 
From Eq.(\ref{w}), the velocity   $W(t_{\mathrm{in}})$ of a star born at $t_{\mathrm{in}}$,
when crossing the plane  ($z \, = \, 0$) is given by
\begin{equation}
W(t_{\mathrm{in}}) \,  \sim  \,  \frac{1}{t_{\mathrm{in}}} \, , \quad \mathrm{thus}  
\quad W(t_{\mathrm{in}}) \, = \, W(t_0) \, \frac{ t_0}{t_{\mathrm{in}}} \, .
\label{W}
\end{equation}
\noindent
This applies at all times, and in particular at present.
We may consider that the trend for the velocity dispersions follows that of the velocities.
   In agrement with the data
   by \citet{Seabroke07} shown in Fig. \ref{dispersion}, we take a value of 10 km s$^{-1}$
    for the present  velocity dispersion $\sigma_{\mathrm{W}}$.
    Thus, as an example  for a group of stars with a mean  age 
   of 10 Gyr, for an age of the universe of 13.8 Gyr the velocity dispersion is 
    estimated to be about  10  km s$^{-1} \,\times \frac{13.8}{3.8} = 36.3$ km s$^{-1}$.
   
   Fig. \ref{dispersion} compares  the corresponding  model predictions obtained in this way (continuous red curve) with the data from \citet{Seabroke07}. 
   We see that the theoretical curve 
   well corresponds to the trend  shown by the observations. We notice that in this plot two different  sources of
   ages are intervening, on one side the ages from stellar evolution and on the other side the cosmic time $t$ 
   intervening in (\ref{W}). Despite this fact, the agreement is quite good and the  growth of the velocity dispersion $\sigma_W$
   for the oldest stellar groups  is well reproduced.
  %The figure suggests that either slightly lower ages of the universe  or vice versa slighlty greater stellar ages would improve
  % the correspondence. 
  In what precedes we have not accounted for the fact  that a galaxy as massive as the Milky Way 
     only forms when the universe is  about 400 millions  years old \citep{Naoz06}. Accounting  for this delay in the star  formation
     leads to   the red broken line in Fig. \ref{dispersion}, which even improves the overall agreement.

    Not only  the flat rotation curves of galaxies, which have been a strong argument in favor of 
     dark matter, appear to be accounted for by the scale invariant equivalent to Newton's law,
     but also the growth  of the ''vertical'' velocity dispersion  with the ages of the stellar groups  in the Galaxy. 
     This result appears consistent with the modified form
     of the Newton's law, derived from the hypothesis of the scale invariance of the macroscopic empty space.
     This does not prove it is right, but at least it shows the interest to pursue this kind of studies.



\end{document}